\documentclass[12pt]{article}
\pdfoutput=1
\usepackage{ifpdf}
\usepackage[a4paper,top=3.2cm,width=17cm, bottom=3.2cm, left=2cm]{geometry}
\usepackage{graphicx}
\usepackage{microtype}
\usepackage[utf8x]{inputenc}
\usepackage[T1]{fontenc}
\usepackage{ae}
\usepackage{aecompl}

\def\bfseries{\fontseries \bfdefault \selectfont \boldmath}

\usepackage[english]{babel}
\usepackage{datetime}
\shortdate 
\usepackage{bm}
\usepackage{subfigure}
\usepackage{amsmath}

\usepackage[normalem]{ulem} 
\usepackage{soul} 
\usepackage[titletoc,title]{appendix}
\usepackage{authblk}
\usepackage{xcolor,soul}
\usepackage[colorlinks=true,linktocpage=true,linkcolor=blue,citecolor=blue]{hyperref}

\usepackage{todonotes}

\usepackage{setspace}

\newcommand{\rd}{\partial}
\newcommand{\wt}{\widetilde}

\usepackage{titlesec}

\titleformat*{\section}{\large\bfseries}
\titleformat*{\subsection}{\bfseries}
\titleformat*{\subsubsection}{\bfseries}

\newcommand{\be}{\begin{equation}}
\newcommand{\ee}{\end{equation}}
\newcommand{\bea}{\begin{eqnarray}}
\newcommand{\eea}{\end{eqnarray}}
\newcommand{\nn}{\nonumber}

\newcommand{\eq}{\begin{equation}}
\newcommand{\eqx}{\end{equation}}

\newcommand{\f}[2]{\frac{#1}{#2}}

\newcommand{\ackno}{\section*{Acknowledgements}}

\begin{document}

\title{{\bf Dynamics near a first order phase transition}}

\renewcommand\Authfont{\scshape\small}

\author[1]{Loredana Bellantuono}
\author[2]{Romuald A. Janik}
\author[3]{Jakub Jankowski}
\author[4,5]{Hesam Soltanpanahi}

\affil[1]{Dipartimento Interateneo di Fisica, Universit\`a degli Studi di Bari, via G. Amendola 173, I-70126 Bari, Italy}
\affil[2]{Institute of Physics, Jagiellonian University, ul. \L{}ojasiewicza 11, 30-348  Krak\'{o}w, Poland}
\affil[3]{Faculty of Physics, University of Warsaw, ul. Pasteura 5, 02-093 Warsaw, Poland}
\affil[4]{Institute of Quantum Matter, School of Physics and Telecommunication Engineering,
South China Normal University, Guangzhou 510006, China}
\affil[5]{School of Physics, Institute for Research in Fundamental Sciences (IPM), P.O.Box 19395-5531, Teheran, Iran}

\date{}
\maketitle
\thispagestyle{empty}

\begin{abstract}

We study various dynamical aspects of systems possessing a first order phase transition in their phase diagram. We isolate three qualitatively distinct types of theories depending on the structure of instabilities
and the nature of the low temperature phase.
The non-equilibrium dynamics is modeled by a dual gravitational theory in 3+1 dimension which is coupled to massive scalar field with self interacting potential. 
By numerically solving the Einstein-matter equations of motion
with various initial configurations, we investigate the structure of the final state arising through
coalescence of phase domains. We find that static phase domains, even quite narrow are very long lived
and we find a phenomenological equation for their lifetime. 
Within our framework we also analyze moving phase domains and their collision
as well as the effects of spinodal instability and dynamical instability on an expanding boost invariant plasma.
\end{abstract}
\newpage
\tableofcontents

\section{Introduction}

Although theories with phase transitions were studied since the very early
days of the AdS/CFT correspondence \cite{Witten:1998zw}, their study was largely restricted 
to the equilibrium setting, where they were identified with an appropriate
Hawking-Page transition between two distinct dual holographic spacetimes.
Such a formulation made it particularly intriguing to investigate
the holographic description of the dynamics of a phase transition occuring
in real time. This is not an academic question as the physics of hadronization
in heavy ion collisions is approximately\footnote{Strictly speaking we then have
a sharp crossover \cite{Bazavov:2014pvz}.} understood as a passage from an expanding and cooling
deconfined quark-gluon plasma to a gas of hadrons in the confined phase of QCD.

The question is especially acute, as a passage through a phase transition
would involve a form of interpolation between different spacetimes, and it is 
\emph{a-priori} far from clear whether one can describe such a process
within classical gravity alone. In some cases this can be done, while in others,
as we argue in the present paper, one would need to develop new approaches.

Plasma dynamics has been extensively investigated within the AdS/CFT framework, starting from an anisotropic and far-from-equilibrium state, analogous to the one produced in heavy ion collisions. In particular, the evolution of the initial state towards the viscous hydrodynamic regime can be characterized by studying the bulk geometry, and horizon formation process \cite{Chesler:2009cy,Heller:2011ju,Bellantuono:2015hxa,Bellantuono:2016tkh}. Those studies were done mostly within the conformal theory.

Dynamical effects in context of holographic phase transitions
were first investigated by quasinormal modes \cite{Janik:2016btb,Janik:2015iry,Gursoy:2013zxa,Rougemont:2018ivt}, revealing rich dynamics, ranging from expected spinodal region up to novel dynamical instabilites (found also in \cite{Gursoy:2016ggq}). A natural step forward was to aim at non-linear time evolution starting from an unstable configuration and following the dynamics of the system up to the final state composed of domains of different phases  \cite{Janik:2017ykj,Attems:2019yqn}. Dynamics of collisions in the presence of phase transitions was studied in Ref.
\cite{Attems:2018gou}, while applicability of second order hydrodynamics in the final, inhomogeneous state was investigated in Ref. \cite{Attems:2017ezz}.

One should point out that holographic theories which exhibit a first order
phase transition, may nevertheless significantly differ between themselves
in certain relevant respects.
In this paper we consider theories which follow three general classes of
equations of states (see Fig.~\ref{fig:eos3classes}), all of which
exhibit a first order phase transition. Class A and B are characterized
by the fact that both the high and low temperature phase are holographically
described by black holes -- thus the~$1^{\rm st}$ order phase transition
occurs between two different kinds of plasma which can coexist at the
transition temperature $T_c$, where the free energies of the two plasmas
coincide. They differ in the kind of modes which become unstable in the spinodal
phase. Class A exhibits just the standard hydrodynamic instability which occurs
for nonzero momenta. This leads to structure formation and spontaneous breaking of
translational invariance. This instability, when followed at the nonlinear level,
leads to the appearance of domains of the two coexisting phases separated
by domain walls. This was numerically demonstrated in \cite{Janik:2017ykj}.
Class B theories exhibit, in addition, a dynamical instability which occurs
even for zero momentum in some range of temperatures within the unstable spinoidal
phase. The unstable mode here is a nonhydrodynamic quasi-normal mode.

The final class of theories, which we denoted by class C, is physically most
interesting. For these theories, black holes exist only up to some minimal temperature,
and therefore the low temperature phase has to be of a thermal gas type (which
means that the background geometry is just the zero temperature one with
Euclidean time compactified). In this case there is no horizon and the low temperature
phase is confining (in the sense of the scaling of entropy with $N_c^0$ instead
of with $N_c^2$). Here the phase transition is a confinement-deconfinement one
and this setup is the most interesting for applications e.g. in heavy ion collisions.

In this paper, we would like to address several open questions concerning the
real time dynamics of theories with a $1^{\rm st}$ order phase transition in the context
of the above three classes of theories. In addition, we provide all the details
of the setup and numerics which were not presented in the initial short paper \cite{Janik:2017ykj}.

The plan of the paper is as follows. In section \ref{sec2}, we will summarize the key
physics questions that we want to address in this paper. In section \ref{sec3}, we start with introducing a general class of holographic models we are interested in this paper. 
We review the basic features of the homogeneous black hole solutions in three classes of the potential and point out their differences in terms of their equation of states.
General frameworks to study the non-equilibrium dynamics of the black holes 
both at linearized and nonlinearized level are also reviewed in this section.
Our results in different setups are presented in sections \ref{sec4}, \ref{sec5} and \ref{sec6}. 
Section \ref{sec4} is devoted to analyzing various aspects of the final states
such as the number of distinct phase domains of the coexisting phases.
We also analyze quantitatively the life time of a narrow domain and the 
coalescence of the neighbouring ones.
Moving domains along the inhomogeneous direction are investigated in section \ref{sec5},
in which, first we give a velocity to a high-energy domain and then we study the collision of two of them moving towards each other. 
In section \ref{sec6}, motivated by realistic heavy-ion collisions, we trace out the effects of first order phase transition and dynamical instability on a boost invariant expanding plasma.  
The potential  which exhibits a confinement-deconfinement phase transition is considered in section \ref{sec7}, where we also comment on the numerical difficulties in this case.
We conclude the paper by a summary and an outlook.  
For completeness appendixes \ref{appA} and \ref{appB} respectively contain some technical details of numerical calculation and holographic renormalization \cite{deHaro:2000vlm}. We adopt a general ansatz which can be directly used also for the boost invariant case.

\section{Main questions}
\label{sec2}

In \cite{Janik:2017ykj}, we demonstrated within the holographic description, that starting from the unstable branch and
adding a small perturbation leads to the formation of domains of the two coexisiting phases as expected physically. While the final state has inhomogeneous horizon/energy, it has uniform free energy\footnote{Apart from the locations of the domain walls.} and Hawking temperature equals to the critical temperature, $T_c$. In the concrete numerical simulations we observed
a single domain of the high energy phase and a single one of the low energy phase.

\subsubsection*{Question 1. What is the generic final state starting from the spinodal branch? Can we describe collisions and coalescence of phase domains?}

It is interesting to ask what would be the final state if one starts from a generic
perturbation with several separated maxima. Would one get at the end several
domains or, on the other hand, would these domains eventually collide\footnote{Similar physics is very relevant in 3+1 dimensions in a cosmological context, see e.g. \cite{Hawking:1982ga,Dymnikova:2000dy}.}
 and coalesce 
to form just two domains. The latter outcome would be preferable (thermodynamically) 
as it minimizes the number of domain walls which increase the free energy.
On the other hand, many domains would lead to interesting black hole solutions.
There is also an intermediate possibility where the well separated multiple domains
would be metastable and exist for a very long time. 
We would like to explore numerically which scenario is realized and whether
we observe collisions and coalescence of well formed phase domains.
We would also like to quantitatively investigate the possible metastability
of configurations with multiple phase domains. 

\subsubsection*{Question 2. What is the impact of a phase transition on boost-invariant expansion?}

Boost-invariant evolution is arguably the simplest setup where we can study a
physical system spontaneously passing through a phase transition. The plasma
system starts off at some high temperature, and then due to boost invariant expansion,
the energy density decreases and the system has to pass through the region
of phase transitions. It is interesting to verify, whether the instabilities
observed in the spinodal phase modify the evolution in a significant way.

\subsubsection*{Question 3. What is the impact of the dynamical instability?}

Holographic models of class B, posses an additional nonhydrodynamic unstable mode
which appears in some subregion of the spinodal phase. This mode is very characteristic
as the instability occurs even for zero momentum, thus leading to an instability
even in the homogeneous case. We would like to see what are the differences in
the dynamics corresponding to the presence of this mode.

\subsubsection*{Question 4. Can we see the confinement-deconfinement phase transition
in real time (holographic) evolution?}

The separation of phases between a black hole phase and the thermal gas phase is
numerically extremely difficult to observe within a single numerical simulation
due to the very different topologies of the two geometries. We adopted here a less
ambitious goal and decided to follow numerical evolution within a black hole ansatz
and check whether we see a \emph{consistent} breakdown of the simulation due 
e.g. regions of high curvature appearing in the numerical domain etc.

\section{Equilibrium and time dependent formulation}
\label{sec3}

In this section we will review the general class of holographic models
that we consider in the present paper.
We concentrate on 3-dimensional theories with 4-dimensional bulk dual,
as in this case we avoid logarithmic terms in the near boundary expansion of the geometry which would severely complicate the numerics.

\subsection{Holographic models and equations of state}

Following bottom-up approach of 
Ref. \cite{Gubser:2008ny,Gursoy:2007er,Gursoy:2007cb}
we use Einstein's gravity coupled to a real scalar field
governed by an action 
\be
S=\frac{1}{2\kappa_4^2}\int d^4x \sqrt{g}  \left[ R-\frac{1}{2}\, \left( \partial \phi \right)^2 - V(\phi) \, \right]~,
\label{Action}
\ee 
where $V(\phi)$ is thus far arbitrary and $\kappa_4$ is related to  four dimensional Newton constant by $\kappa_4=\sqrt{8\pi G_4}$.

Since we are interested in asymptotically $AdS$ space-time geometry, 
the potential needs to have the following small $\phi$ expansion
\be
V(\phi)\sim -\frac{6}{L^2} + \frac{1}{2}m^2\phi^2 + O(\phi^4)~. 
\ee
Here, $L$ is the $AdS$ radius, which we set it to one, $L=1$, by the 
freedom of the choice of units.
Such a gravity dual corresponds to relevant deformations
of the boundary conformal field theory
\be
\mathcal{L} = \mathcal{L}_{\rm CFT} + \Lambda^{3-\Delta} O_{\phi}~,
\ee
where $\Lambda$ is an energy scale, and $\Delta$ is a conformal
dimension of the operator $O_\phi$ related to the mass parameter of the scalar field according to holography, $\Delta(\Delta-3)=m^2$. 
We consider $3/2\leq\Delta<3$ which corresponds to relevant 
deformations of the CFT and respects the Breitenlohner- Freedman bound, $m^2\geq-9/4$ \cite{Breitenlohner:1982bm,Breitenlohner:1982jf}.

To find the phase structure we solve Einstein's equations coupled
to the scalar field with $AdS$ boundary conditions. It is convenient
to use the following coordinate system
\be
ds^2= e^{2 A(r)}\left(-h(r)dt^2 + d\vec{x}^2\right) - 2e^{A(r)+B(r)} dr dt~,
\ee
with $\phi(r)=r$ gauge \cite{Gubser:2008ny}. A static event horizon requires a condition $h(r_H)=0$ for some $r_H$.

Entropy density and temperature are readily obtained
from the horizon area and regularity of the space time     respectively,
\be
s= \frac{2\pi}{\kappa_4^2}e^{2 A(r_H)}, \hspace{50pt} T=\frac{e^{A(r_H)+B(r_H)}|V'(r_H)|}{4 \pi}~.
\ee
Physical quantities of the boundary theory are read off from the geometric data in a standard way
by means of holographic renormalization \cite{deHaro:2000vlm,Elvang:2016tzz,Skenderis:2002wp},
while the speed of sound of the system could be computed via either horizon or boundary data as
\be
c_s^2=\frac{d \ln(T)}{d \ln(s)}=\frac{d P}{d \epsilon}~.
\ee
The free energy of the system is just the on-shell value of the
action $F = T S_{\rm on-shell}$. 
Since we are using minimal terms in holographic renormalization procedure to cancel the divergencies, it is more convenient to use the relation between free energy, entropy and temperature,
\be
\delta F=-\int_{T_0}^{T} s(\tilde{T}) \,d\tilde{T}
\label{eq:FandS}
\ee
where $T_0$ corresponds to the black hole with vanishing entropy (thermal gas).


\begin{figure}
	\begin{center}
	\includegraphics[height=.20\textheight]{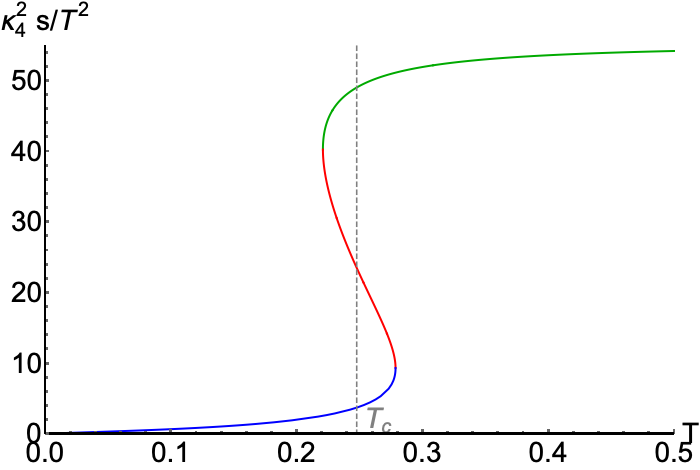} \hspace{10pt}
    \includegraphics[height=.20\textheight]{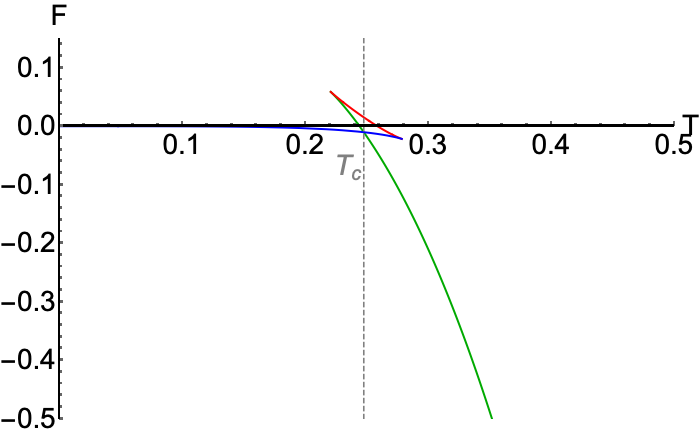}\\
    \vspace{15pt}
    \includegraphics[height=.20\textheight]{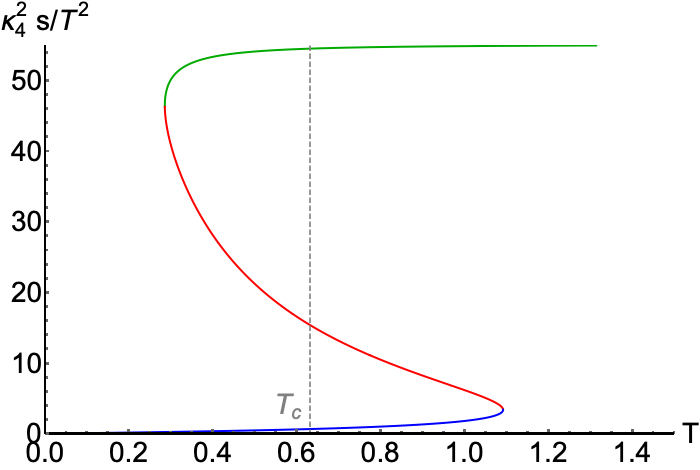}\hspace{10pt}
    \includegraphics[height=.20\textheight]{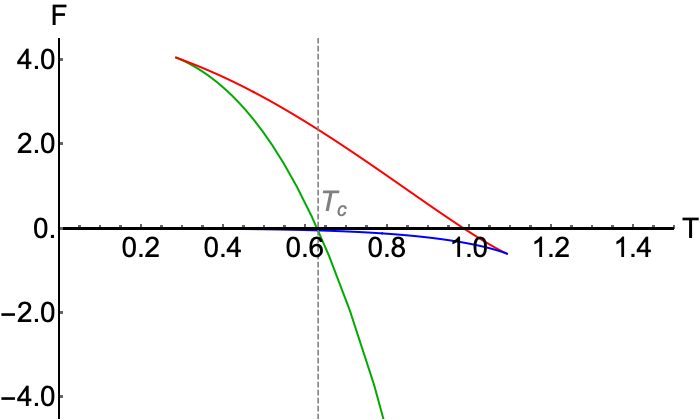}\\
    \vspace{15pt}
    \includegraphics[height=.20\textheight]{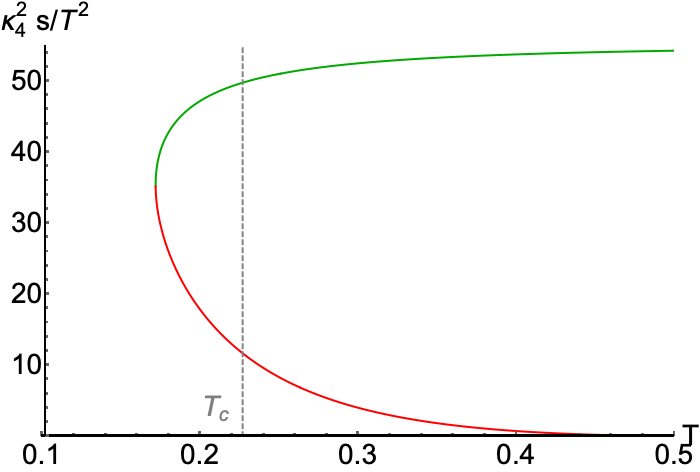}\hspace{10pt}
    \includegraphics[height=.20\textheight]{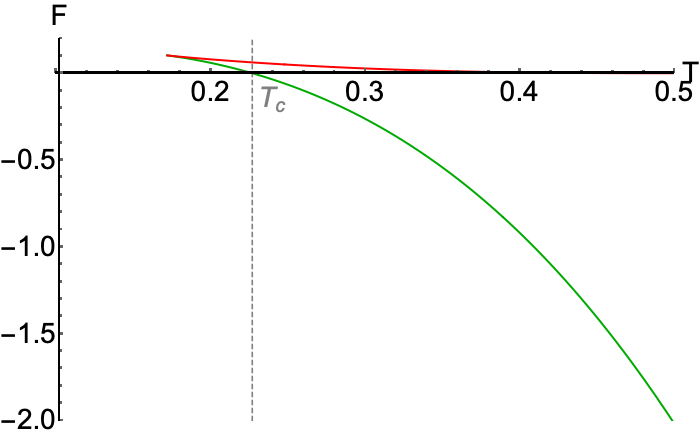}\\
\caption{Equations of state for potentials of classes A (top panels), B (middle panels), C (bottom panels). The Bekenstein-Hawking entropy density (left panels) are given by the horizon area and the free energy (right panels) are computed using the thermodynamic relation of Eq.~(\ref{eq:FandS}). 
Green and blue lines represent respectively high and low temperature stable states, while red segments correspond to unstable regions.}
 \label{fig:eos3classes}
\end{center}	
\end{figure}


\begin{table}[h]
\begin{center}
 \begin{tabular}{||c c c c||} 
 \hline
 $V(\phi)$ & Class A & Class B & Class C \\ [0.5ex] 
 \hline
 \hline
 $\gamma$ & $1/\sqrt{3}$ & $1/\sqrt{3}$ & 1\\ 
 \hline
 $b_2$ & 0 & 0 & 2 \\
 \hline
 $b_4$ & -0.2 & -0.3 & 0.1 \\  [1ex] 
 \hline
\end{tabular}
\caption{Parameters for three classes of used potentials.}
\label{tabV}
\end{center}
\end{table}


Introducing different potentials for the scalar field may lead to different phase structures. To be more precise let us consider
the following parametrization of the scalar self-interaction potential
\bea
V(\phi)=-6\,\cosh\left(\gamma \phi\right)
+b_2\,\phi^2 + b_4\,\phi^4~,
\label{V}
\eea
and define three classes of parameter sets leading to different
equations of state as summarized in Table \ref{tabV} and in Fig.
\ref{fig:eos3classes}.
In all cases the resulting equations of state
exhibit a first order phase transition and the critical temperature $T_c$ for each case  can be found by computing the free energy of the black holes and seeking for the solution with lowest free energy at a given temperature.
Classes A and B are characterized
by the fact that both the high and low temperature phases are holographically
described by black holes -- thus the $1^{\rm st}$ order phase transition
occurs between two different kinds of plasma, which can coexist at the
transition temperature $T_c$, where the free energies of the two plasmas
coincide. They differ in the kind of modes which become unstable in the spinodal
phase, which we will review shortly.

As was already described in the introduction, the physically most interesting class of theories is denoted by class C.
As we mentioned, there is a minimum temperature below which  black hole solutions do not exist. In other words, the only solution for low temperature is a thermal gas configuration, which is a geometry without event horizon, and exists at any temperature with compactified Euclidean time.
While the black hole can resemble the deconfined phase of the dual theory, the thermal gas corresponds to the confined one.
Therefore, in this case, the first order phase transition is a confinement-deconfinement one, and this setup is the most interesting for applications e.g. in heavy ion collisions.

\subsection{Linear perturbations and stability regions}

In order to study the response of the system to small
perturbations consider
\be
g_{ab}=g_{ab}^{\rm BH}(r)+ h_{ab}(r)e^{i kx -i \omega t},~\hspace{20pt}
  		\Phi=\Phi^{\rm BH}(r)+ \phi(r)e^{i kx -i \omega t} ~,
\ee
where the BH subscript refers to the background metric of the 
previous subsection. A standard approach is to group linearized
functions into gauge invariant objects resulting in a few coupled
channels describing various effects \cite{Kovtun:2005ev}.

In the sound channel, generically for systems with a first order
phase transition there exists a {\it spinodal} instability, which is
characterized by the negative value of the square of the speed of sound.
In turn the sound mode has the following dispersion relation
\be
\omega\approx\pm\, i |c_s|\, k - \,\frac{i}{2\,T}\,\left(\frac{4}{3}\,\frac{\eta}{s}+\frac{\zeta}{s}\right)\, k^2=
		 \pm i |c_s|\, k - i\Gamma_s k^2~,
\ee
where $c_s$ is the speed of sound, $k$ is the momentum, $T$ is the Hawking temperature, $s$ is the entropy density, and $\eta, \zeta$ are respectively shear and bulk viscosities.
It is easy to see that for small enough $k$ we have ${\rm Im}\ \omega>0$.
This mode is only present for a finite range of momenta $0<k<k_{\rm max}$
with $k_{\rm max}=|c_s|/\Gamma_s$. The maximal value of $\omega$ in this
range is called the growth rate.

We can return now to our main classes of theories and specify the difference
between class A and B.
Class A exhibits just the above standard hydrodynamic instability which occurs
for nonzero momenta. This leads to structure formation and spontaneous breaking of
translational invariance. This instability, when followed at the nonlinear level,
leads to the appearance of domains of the two coexisting phases separated
by domain walls. This was numerically demonstrated in \cite{Janik:2017ykj}.
Class B theories exhibit, in addition, a dynamical instability which occurs
even for zero momentum in some range of temperatures within the unstable spinodal
phase \cite{Janik:2016btb,Gursoy:2016ggq}. The unstable mode here is a nonhydrodynamic quasi-normal mode.

The full study of the QNM spectrum for all three classes of potential has been done in Ref. \cite{Janik:2016btb,Janik:2017ykj} for the higher dimensional analogues. In the $AdS_4/CFT_3$ case we also find similar behaviour, and as advocated in the previous section, on top of spinodally unstable region class B of potentials contains a dynamical instability through a non-hydrodynamic mode.

\subsection{Time dependent geometries and nonlinear evolution}

The most interesting questions concerning real time dynamics
of theories with a $1^{\rm st}$ order phase transition remain in
the realm of nonlinear evolution. Similarly to many other studies in numerical holography \cite{Chesler:2008hg,Chesler:2013lia},
it is convenient to use the Eddington-Finkelstein coordinate system (EF).
Our concrete parametrization of the metric is given by
\bea
\label{eq:metrictimedep}
ds^2=-A\,dv^2-\frac{2\,dv\,dz}{z^2}-2\,B\, dv\,dx + S^2\,\left( G\,dx^2+G^{-1}\, dy^2 \right)
\eea
where $A, B, S, G$ and $\phi$ are functions of $(z, v, x)$, where
$v$ is the Eddington-Finkelstein time, and $z$ is the holographic coordinate. 
On the boundary $z=0$, the Eddington-Finkelstein time $v$ coincides
with the conventional Minkowski time $t$ (or in the boost-invariant case
considered later in the paper, with the longitudinal proper time $\tau$).
Hence, in all our plots we will use the conventional Minkowski notation
$t$ or $\tau$.

In appendix~\ref{appA}, we provide the details on the numerical procedure
for carrying out time evolution adopted in the present paper.
The initial conditions for the
evolution are given by specifying the initial profiles of the
functions $S(z,v_0,x)$, $G(z,v_0,x)$ and the 
leading boundary asymptotics of $B$. See appendix~\ref{appA} for the details.

Performing holographic renormalization, we extract the physical observables of interest -- the components of the energy-momentum tensor
as well as the expectation value of the operator dual to the
bulk scalar field. These observables are given  
through the near boundary expansion of the metric coefficients and the scalar field. We sketch the derivation and provide explicit formulas in appendix~\ref{appB}. 

\section{Universality aspects of the final state}
\label{sec4}

In this section we will study results of time evolution
with various perturbations on top of the spinodal regime. 
These perturbations will trigger the spinodal instabilities and will develop
further following fully nonlinear evolution. 
The focus will be on universality aspects of the final state.
In the previous paper \cite{Janik:2017ykj}, we showed that the system
undergoes phase separation and two domains of the two coexisting phases (with
equal free energy) are formed, separated by domain walls.

A natural question, as indicated in the introduction, is what happens when we have well separated perturbations. Will multiple domains form, or will they eventually coalesce into a single domain of each phase, thus minimizing the number of domain walls? What is the time scale of this dynamics?

\begin{figure}
\begin{center}
\includegraphics[height = .27\textwidth]{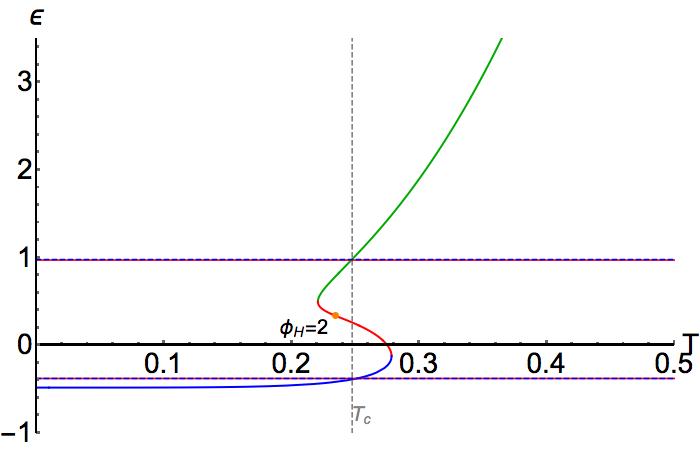}\hspace{10pt}
\includegraphics[height = .27\textwidth]{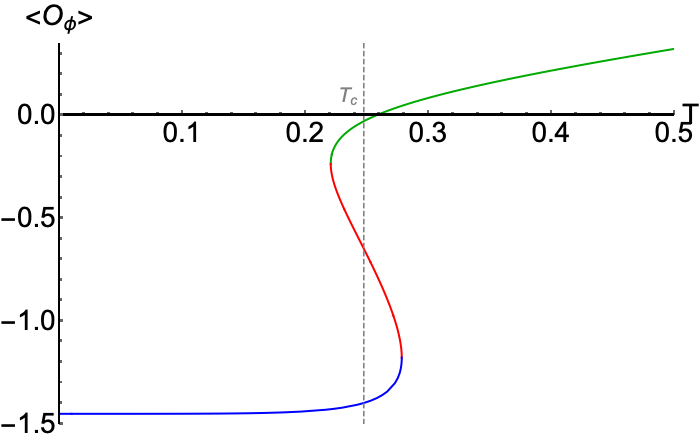}
\caption{Energy density (left panel) and the expectation value of the operator dual to the bulk scalar field (right panel) for a system of class A \cite{Janik:2017ykj}. The indicated point $\phi_H=2$ is used as a initial configuration for time evolution.}
\label{fig:markedEOS}
\end{center}
\end{figure}

In this section, we perform simulations of the same system of class A
as in the previous paper, but with larger spatial domains and initial conditions with well separated perturbations. For completeness, we recall the plot of the energy density
in Fig.~\ref{fig:markedEOS}. We will refer to the two phases marked by horizontal lines
as the low and high energy phase (which coexist at $T_c$).


\subsection{General considerations on phase domain formation}

We will now explore a family of initial 
conditions starting from the unstable spinodal branch
perturbed by two bumps localized in two regions of the
spatial domain. We expect that the instabilities seeded
by the bumps will grow and develop into several domains
of the coexisting phases. The aim is to see whether
these multiple domains will persist or whether the domains
will coalesce and lead to a final state with just a single domain
of each of the two phases.

As initial configuration we start with a static homogeneous black hole solution in unstable branch with $\phi_H=2$ and we add a perturbation in $S$ function. The detailed shape of perturbing function is
\begin{equation}
    \delta \widetilde{S}(z,x) = \widetilde{S}_0 z^2(1-z)^3\left[\exp\left(-w_0\cos\left(k\left(x-\f{a_B}{2}\right)\right)^2\right)+
    \alpha \exp\left(-w_0\cos\left(k\left(x+\f{a_B}{2}\right)\right)^2\right)\right]~, 
\end{equation}
where the parameter $\alpha$ determines the asymmetry of the configuration.
The $x$ periodicity is $24\pi$ and we set $k=1/24$.
On top of that we choose $\widetilde{S}_0=0.1$, $w_0=5$ and $a_B=15\pi$. 

\begin{figure}
	\begin{center}
	\includegraphics[height=.26\textheight]{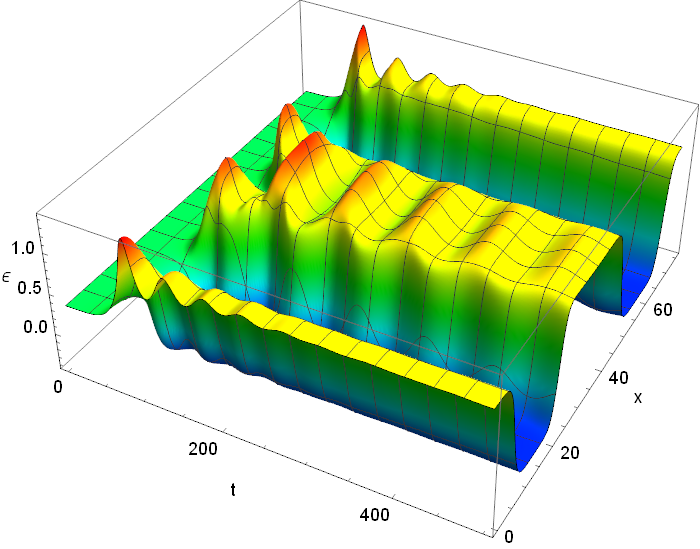}
	\includegraphics[height=.26\textheight]{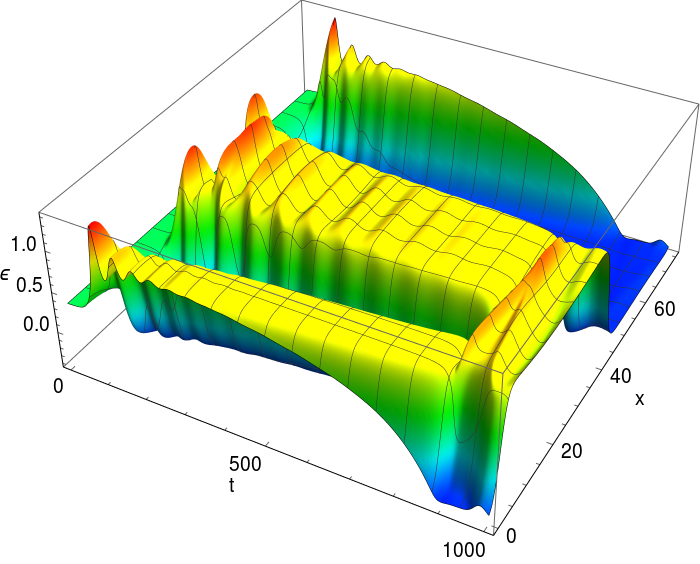}
	\includegraphics[height=.26\textheight]{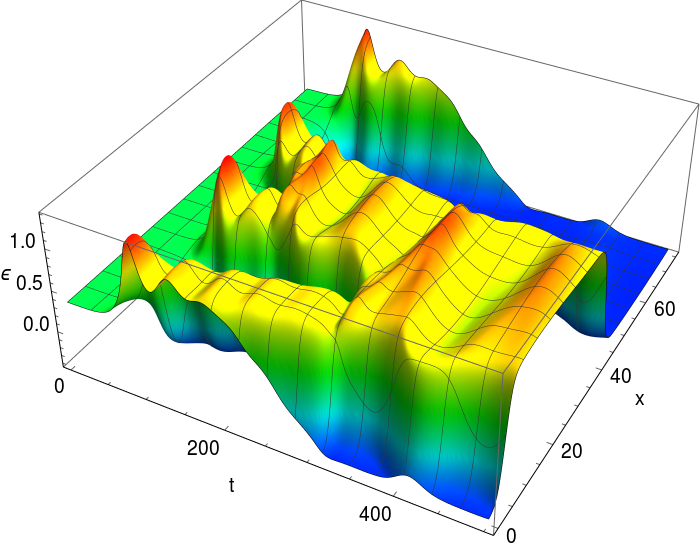}
	\includegraphics[height=.26\textheight]{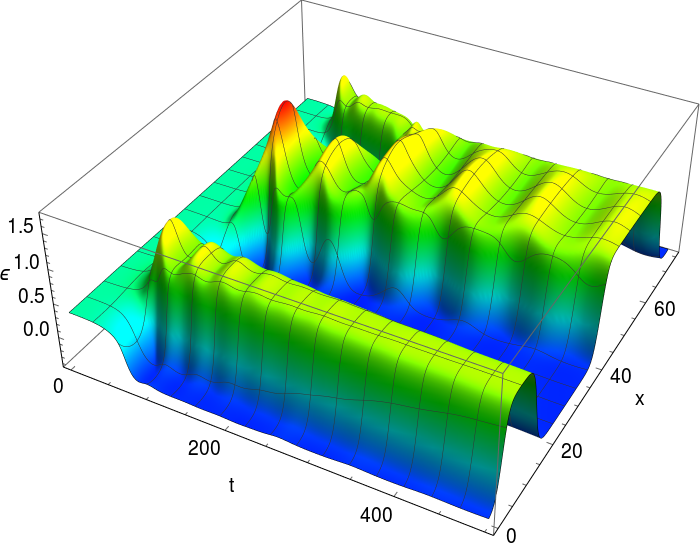}\caption{Time dependence of energy density for initial perturbations with varying degree of asymmetry. Top row: $\alpha=0.5$ (left) and $\alpha=0.25$ (right), bottom row: $\alpha=0.1$ (left) and $\alpha=-1.0$ (right).} \label{fig:space_time_energy}
\end{center}	
\end{figure}

We performed simulations in the symmetric case $\alpha=+1$, and several
cases with increasing asymmetry $\alpha=0.5, 0.25, 0.1, -1$. The temporal
evolution of the energy density is shown in Figure~\ref{fig:space_time_energy}.
The symmetric case ($\alpha=+1$) is visually indistinguishable\footnote{This seems
to occur because the asymmetric component of the perturbation seems to die down
already in the initial linear regime before the nonlinear evolution kicks in. For larger asymmetry $\alpha<0.5$, the asymmetry persists in the nonlinear regime.}
from the case with the lowest asymmetry ($\alpha=0.5$) shown in the top left corner.

In all cases we see initially three domains of the high energy phase around $t\sim 50-100$. Two of these merge relatively quickly forming a longer lived
state with two domains of each phase. Subsequently in all asymmetric cases, apart
from $\alpha=0.5$, those two domains eventually merge leaving just a single
domain of each phase. Note, however, that that process may be very slow (see e.g.
$\alpha=0.25$ in the top right corner). Indeed, one may speculate that this merging could also occur for
$\alpha=0.5$ but at a time scale at least of order of magnitude longer.
We extended that simulation beyond $t=500$ but did not observe \emph{any} decrease in 
the distance between the two high energy domains which would have to occur prior to merging. We also checked that perturbing the domain wall did not change the
behaviour. Thus the observed meta-stability for well separated domains seems
to be quite robust. Of course, for symmetry reasons, in the symmetric case ($\alpha=+1$)
we expect this two domain configuration to be the final one.


\begin{figure}
	\centering
		\includegraphics[width = 0.7\textwidth]{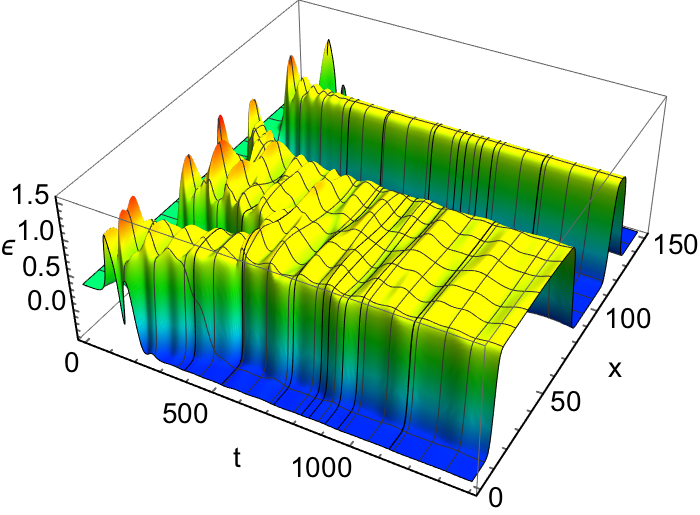}	
	\caption{Space and time dependence of energy density for a configuration with $\alpha=0.1$ displaying eight high temperature domains in the initial evolution.}\label{fig:space_time_energy_2}
\end{figure}


In Fig. \ref{fig:space_time_energy_2} we show energy
density for a time evolution with $\alpha=0.1$ asymetry,
$a_B=30\pi$ and $k=1/48$ in the spatial box of $48\pi$ extent. In this case, at early times,
eight narrow bumps of high temperature phase appear and on a short time scale those merge into three larger domains which subsequently form two high temperature domains of different size. We have checked that from $t\simeq500$ up to $t\simeq2000$ no interesting dynamics appear, suggesting that this state is meta stable. We strongly suspect, that if followed to much larger time scales the system could still change the pattern of domains.

\subsection{Quantitative analysis of merging domains}

Although the results of the simulations presented above exhibit merging of domains,
this can be seen predominantly for filling up quite narrow low energy domains,
or as a result of collisions of domains of the high energy phase which are formed independently and which move towards each other and eventually collide.
Slightly wider domains tend to persist for a long time, in some cases even throughout the duration of our numerical simulations.
In order to quantify the life time of the domains as a function of their width,
we constructed a set of initial conditions, where we can tune the width of one
of the low energy domains, and have at the same time a static initial configuration.

\begin{figure}
	\begin{center}
	\includegraphics[height=.26\textheight]{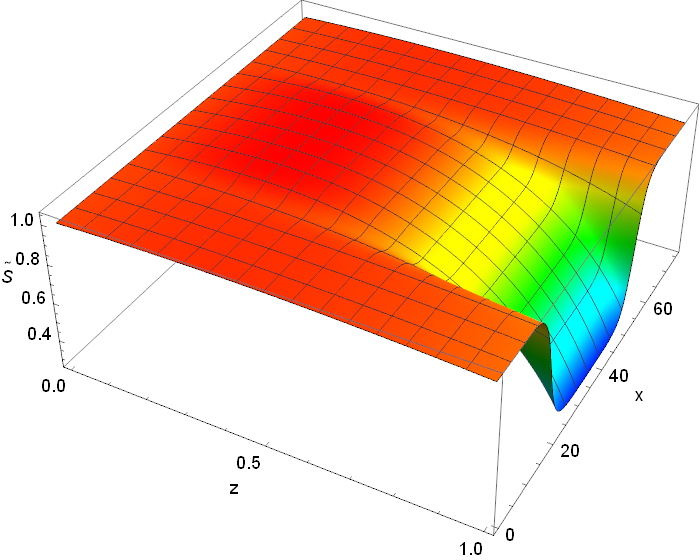}\\
	\includegraphics[height=.26\textheight]{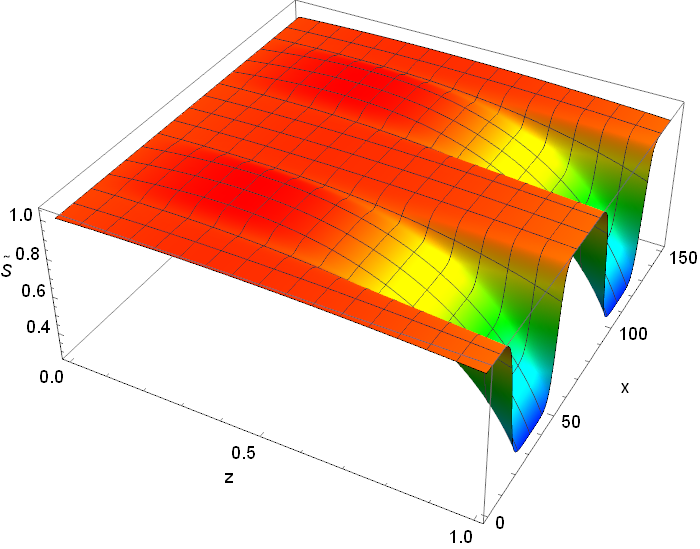}
	\includegraphics[height=.26\textheight]{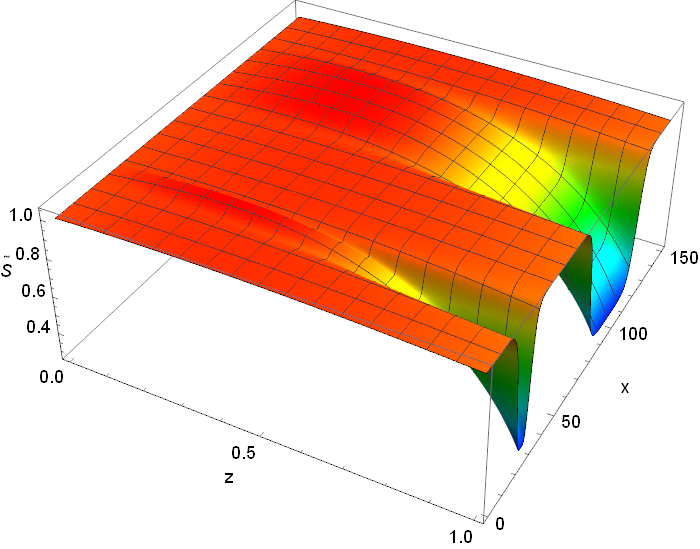}
\caption{Preparing the initial conditions for $\tilde{S}(z,x)$: the original static final state from \cite{Janik:2017ykj} at the top, 
the doubled configuration (bottom left) and the subsequent deformation by the nonlinear $g(x)$ function (\ref{e.g}) (bottom right).\label{fig:squashedinit}}  
\end{center}
\end{figure}

The construction of these initial configurations (for $\tilde{S}(z,x)$) is sketched in Fig.~\ref{fig:squashedinit}. We start
with the static final state\footnote{The one with $\phi_H=2.0$ from that paper.} of \cite{Janik:2017ykj}, then we double the
period, obtaining a configuration with two equal domains each of the high and
low energy phase. Then we construct a new initial configuration by the formulas
\eq
\tilde{S}_{deformed}(z,x) = \tilde{S}(z, g(x))  \quad\quad\quad\quad
\tilde{G}_{deformed}(z,x) = \tilde{G}(z, g(x)) 
\eqx
with the "squashing function" $g(x)$ of the form
\eq
\label{e.g}
g(x) = \alpha x + \beta \tanh \gamma (x-x_0)
\eqx
with appropriately chosen parameters. We set the remaining initial condition $b_1(x)=0$.

\begin{figure}
	\begin{center}
	\includegraphics[height=.3\textheight]{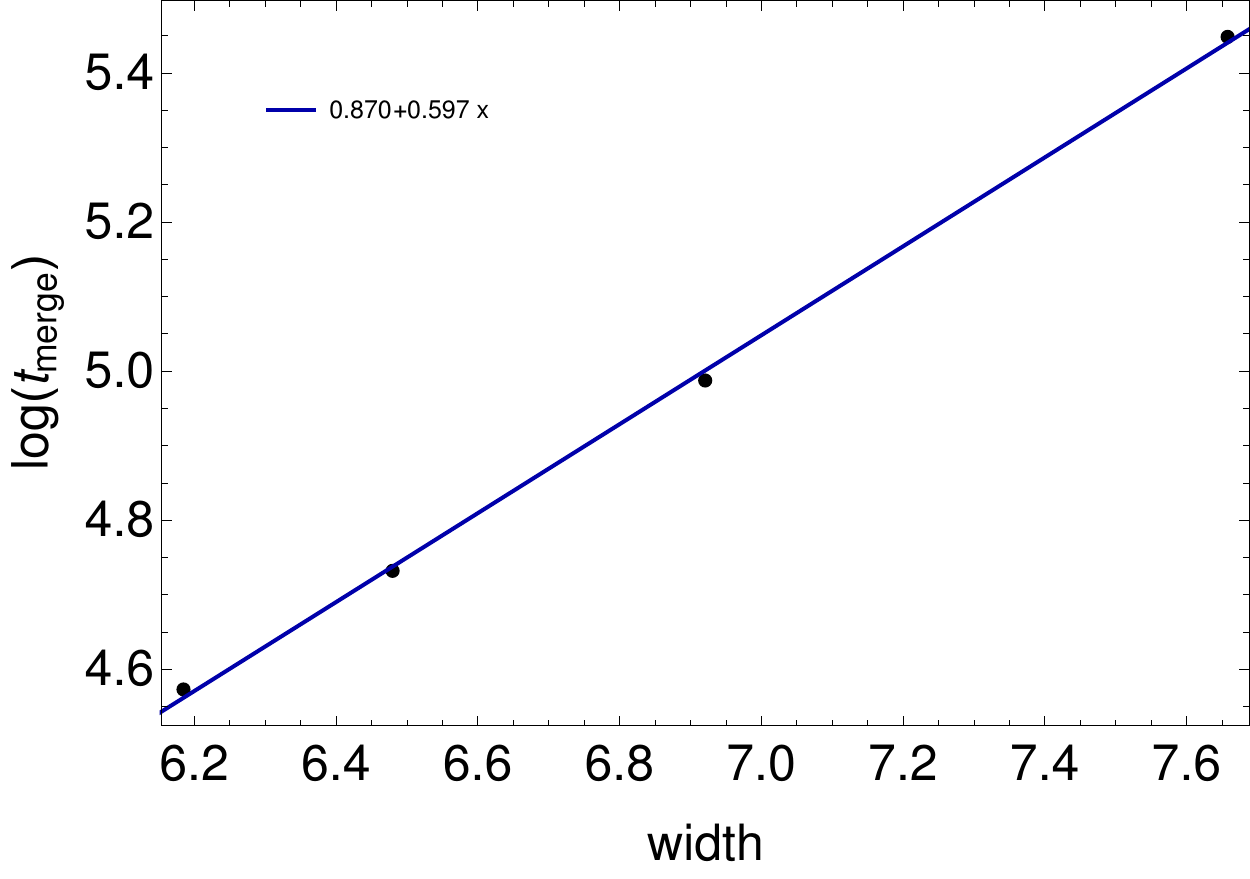}
\caption{Merging time as a function of the width of the domain of the low energy phase.\label{fig:mergetime}}  
\end{center}
\end{figure}

We define the width of the domain as the width of the region with energy $\epsilon<0.5$. The value $0.5$ is just chosen for definiteness.
In order to quantify the merging time $t_{merge}$, we measure the time from the beginning
of the simulation until the energy throughout the region rises above $\epsilon=0.5$.
We find an exponential dependence of $t_{merge}$ on the width:
\eq
\log(t_{merge}) = 0.870+0.597\, width
\eqx
which fits well the results of the simulations as shown in Fig.~\ref{fig:mergetime}.
The exponentially long lifetime of even moderately wide domains means that the thermodynamically favoured configuration with the smallest possible number of two domain walls (and just a single domain of each phase) may in some cases be never realized in practice. The dominant mechanism for merging of domains is rather their relative motion and subsequent collisions (seen in various stages of Figures~\ref{fig:space_time_energy} and~\ref{fig:space_time_energy_2}). We also performed a simulation of the motion and the collision of two fully formed phase domains which we review in the following section.

\section{Moving phase domains}
\label{sec5}

In this section we will study the time evolution of a system with phase separation where the high-energy domain moves along the inhomogeneity direction ($x$). We will also comment on the possible application of this scenario to the analysis of a collision of two such domains moving in opposite directions. We will use the same class A potential as in the previous section.

\subsection{Motion of a single phase domain}

\begin{figure}
	\centering
		\includegraphics[width = 0.7\textwidth]{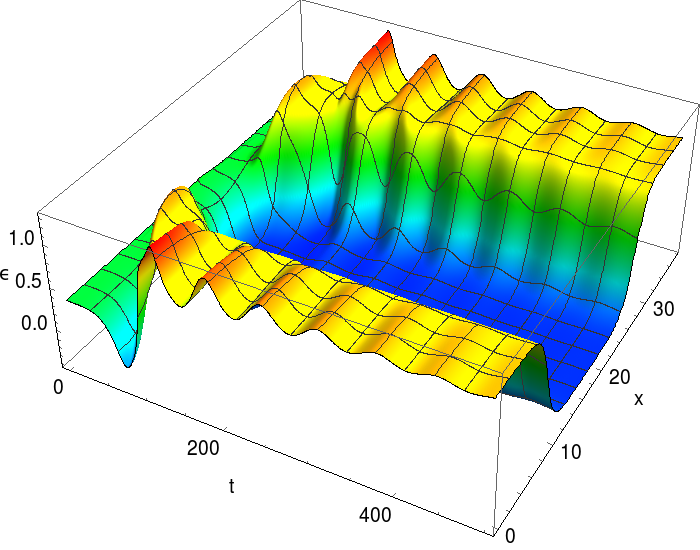}	
	\caption{Space and time dependence of energy density of for $\phi_H=2.0$ with
	 Gaussian perturbation from Ref. \cite{Janik:2017ykj}.}\label{fig:finalenergy}
\end{figure}

To construct our initial conditions, we start from the final state of the simulation in Fig. \ref{fig:finalenergy}, representing a static high energy domain coexisting with a low energy phase. In order to get a moving domain, we modify the static metric functions in that state to have a nonvanishing $\langle T^t_{~x} \rangle$. Specifically, referring to the redefined functions in Eqs. \eqref{asym-inhomo-2}, \eqref{asym-inhomo-4} and \eqref{asym-inhomo-5} in Appendix A, we add a small constant to $\tilde{B}(z,x)$ (namely, we replace $b_1(x) \to b_1(x) + C$), while leaving the functions $\tilde{S}(z,x)$ and $\tilde{G}(z,x)$ unchanged.


\begin{figure}
\begin{center}
\includegraphics[width = 0.45\textwidth]{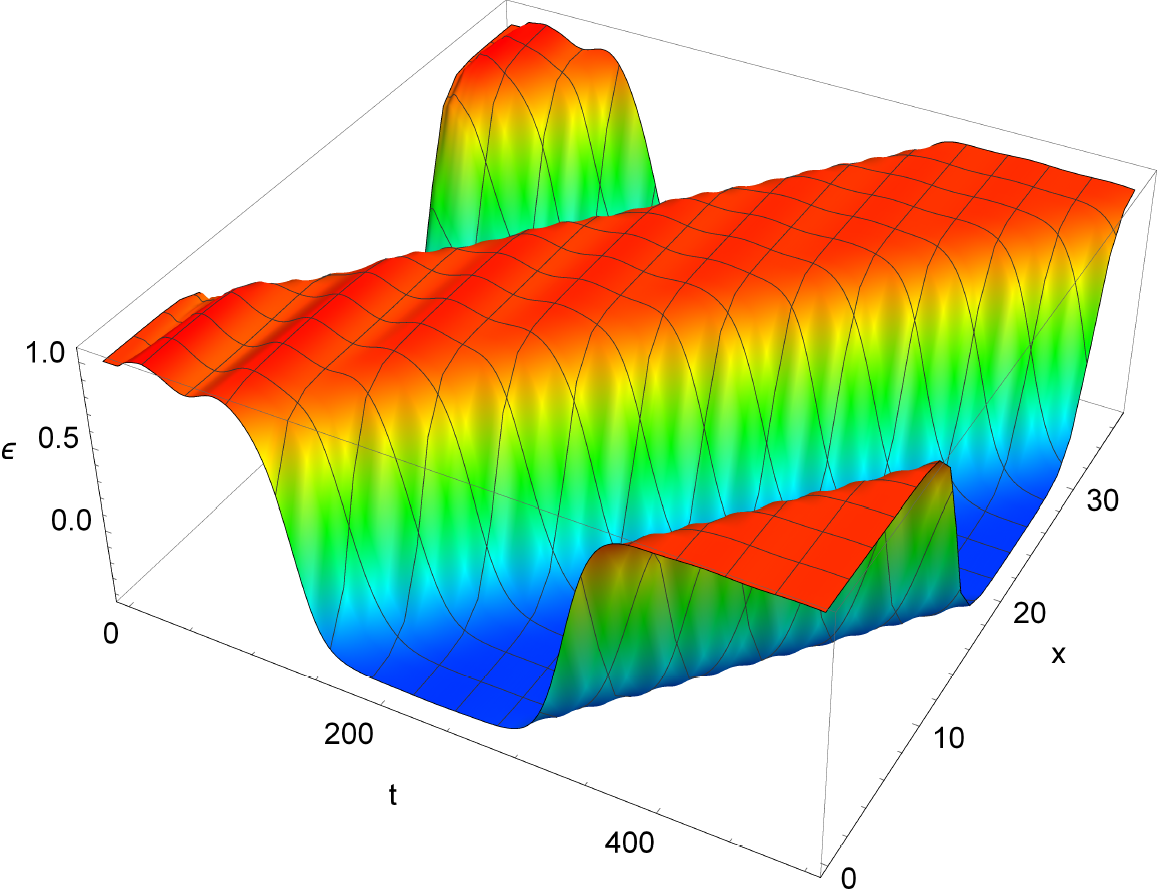}\hspace{0.5cm}
\includegraphics[width = 0.45\textwidth]{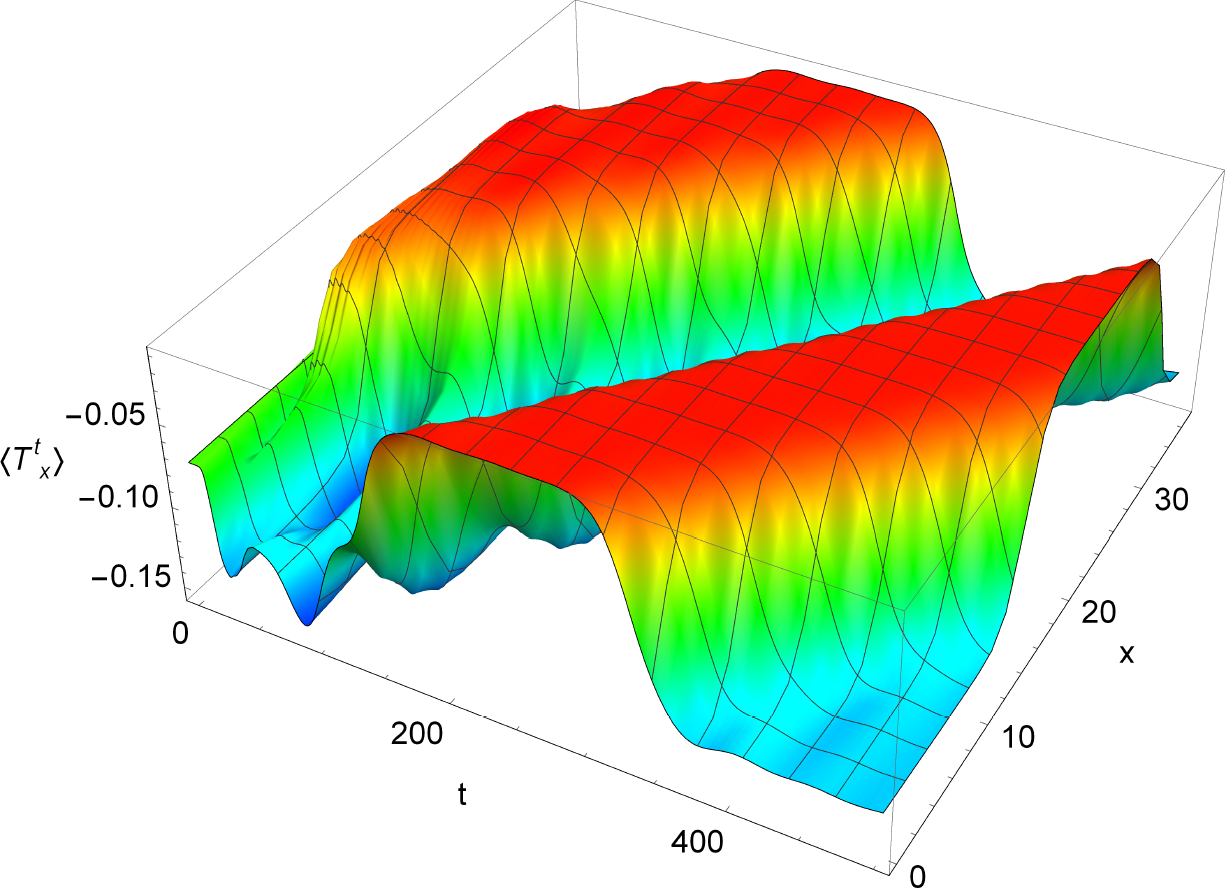}\\
\caption{Space and time dependence of the energy density $\epsilon$ (left) and the transferred momentum density $\langle T^t_{~x} \rangle$ (right). The initial state is obtained from the final snapshot in Fig. \ref{fig:finalenergy}, modified by adding the constant $C=0.05$ to the function $\tilde{B}(z,x)$.}
\label{fig:energy_density_and_Txt_one_domain}
\end{center}
\end{figure}

This leads initially to a nonvanishing practically constant (negative) momentum density $\langle T^t_{~x} \rangle$ throughout the spatial domain. Note, however, that after a short time (see Fig.~\ref{fig:energy_density_and_Txt_one_domain} right), the momentum density for the low energy phase increases practically to zero, while it remains negative for the high energy phase.
Therefore, we obtain a setup of a moving domain of the high energy phase
in the background of a \emph{static} low energy phase. It would be interesting to understand the physical reason for this behaviour. However, this is exactly what we need later for studying collisions of moving domains.


\begin{figure}
	\centering
		\includegraphics[width = 0.60\textwidth]{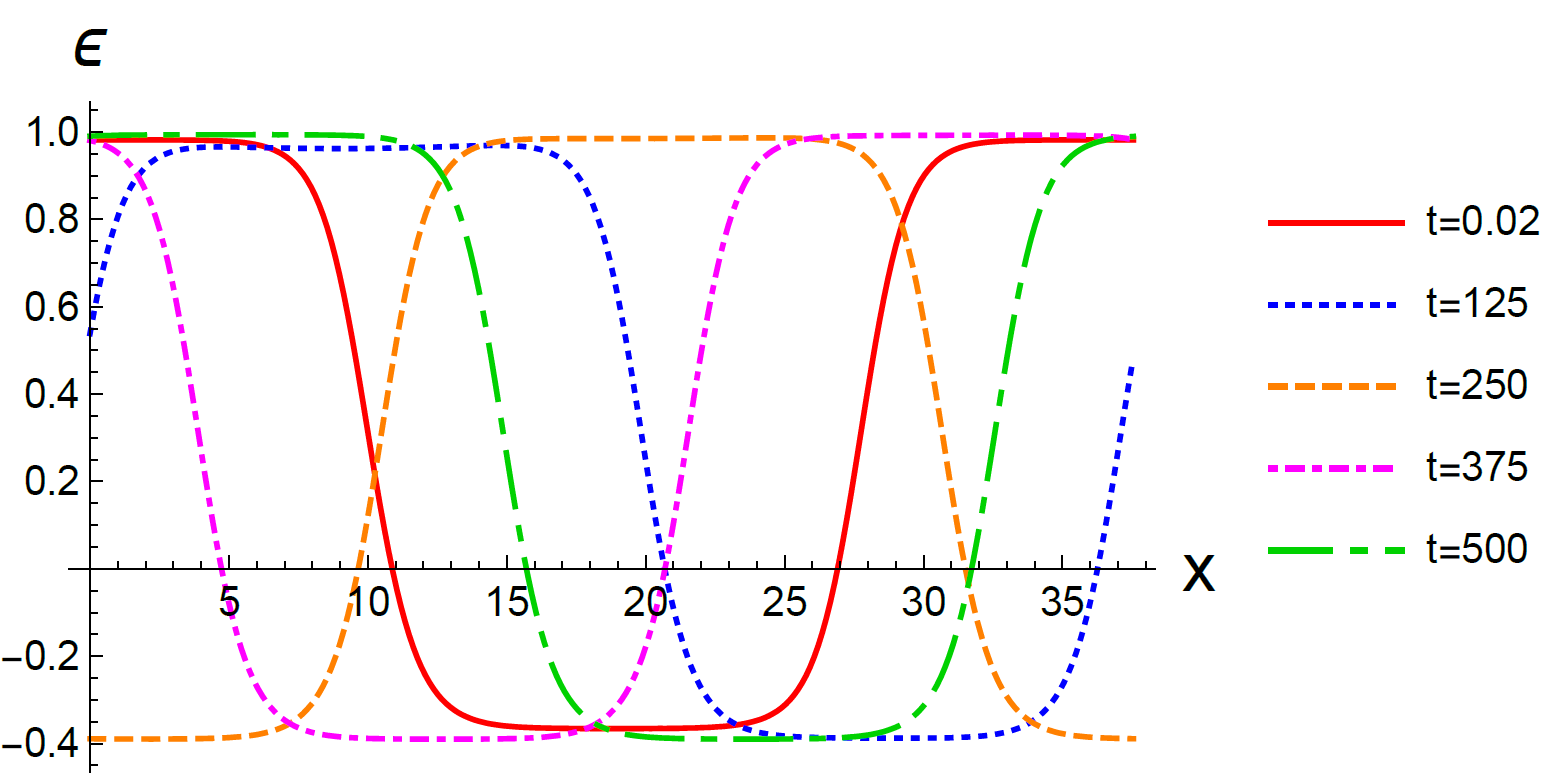}	
	\caption{Evolution of the energy density profile for the initial state obtained from the final snapshot in Fig. \ref{fig:finalenergy}, modified by adding a constant $C=0.05$ to the function $\tilde{B}(z,x)$. 
	}\label{fig:translation_energy_one_domain}
\end{figure}


We solve numerically the Einstein-matter equations associated to the geometry in Eq. \eqref{eq:metrictimedep}, with the aforementioned initial conditions. The results for $C=0.05$ yield a slowly moving domain of the high energy phase on top of a static background with low energy density, as shown in Fig. \ref{fig:translation_energy_one_domain}. 
As this is not a Lorentz boost, we may \emph{a-priori} expect dissipation to occur and observe a gradual slowing down of the high energy domain due to friction from the static low energy phase. In order to check this, we analyzed in detail the movement of the high energy phase.

\begin{figure}
	\centering
		\includegraphics[width = 0.60\textwidth]{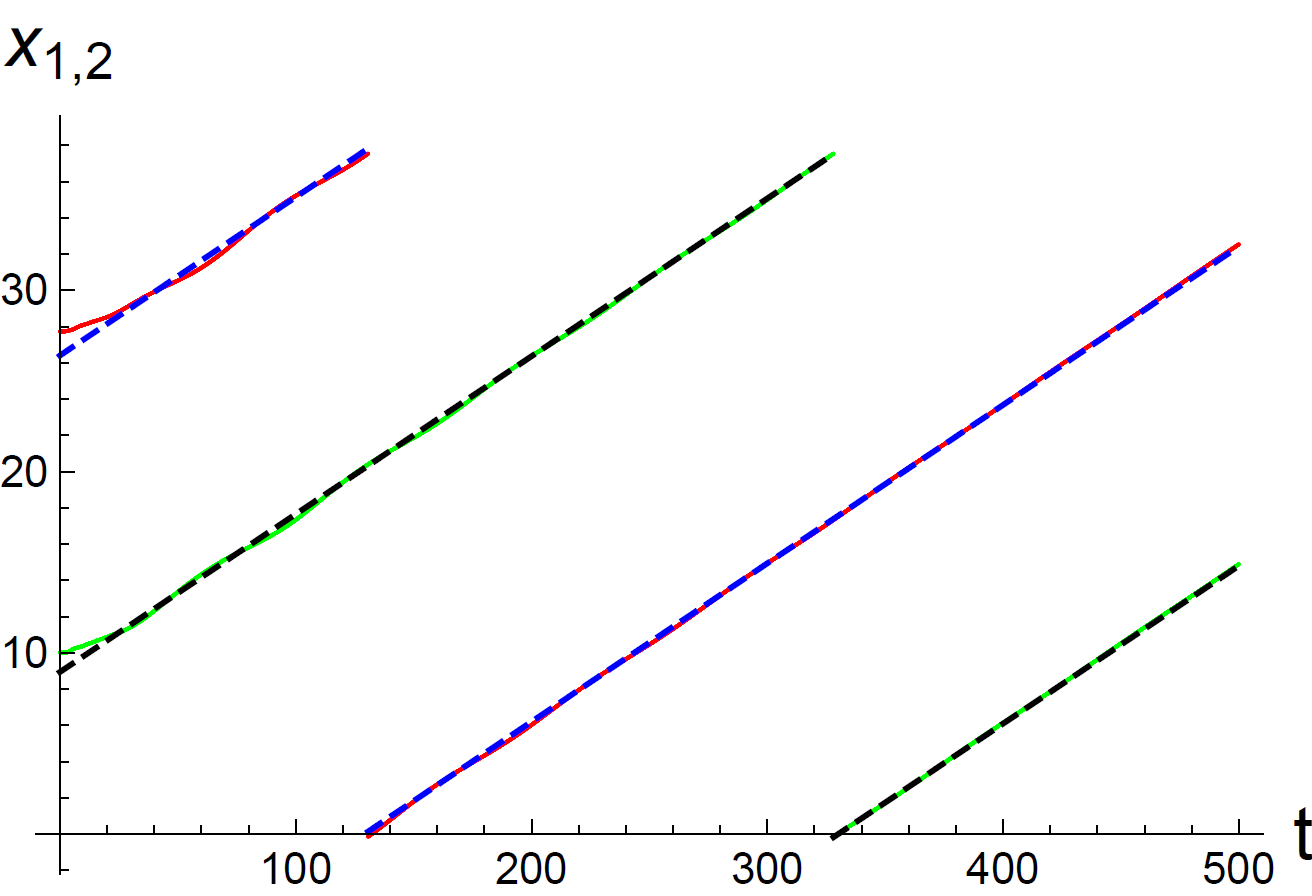}
	\caption{Evolution of the points $x_1(t)$ (red line), on the domain wall with increasing energy, and $x_2(t)$ (green line), on the domain wall with decreasing energy, satisfying $\epsilon(t,x_{1,2}(t))=\left(\max_x \epsilon(t,x)+ \min_x \epsilon(t,x)\right)/2$. The blue and black dashed lines represent the linear fits in Eq.\ \eqref{eq:fit}, with parameters \eqref{eq:fit_x1} and \eqref{eq:fit_x2}, respectively.}\label{fig:fit_one_domain}
\end{figure}

It can be observed that the motion of the domain is approximately a rigid translation along the inhomogeneity direction with constant velocity. However, since during the evolution there is also a slight variation in the energy values of the two phases, the kinematics of the domain has been quantitatively analyzed by following the motion of the points $x_1(t)$ and $x_2(t)$ on the two domain walls, such that $\epsilon(t,x_{1,2}(t))=\left(\max_x \epsilon(t,x)+ \min_x \epsilon(t,x)\right)/2$. By convention, we will denote by $x_1$ the point with increasing energy and by $x_2$ the point with decreasing energy. The results of a linear fit in the $(t,x)$ plane yield
\begin{equation}
x_{1,2}(t)= \left(q_{1,2} + r_{1,2} t\right)\mathrm{mod}\, 12\pi \label{eq:fit}
\end{equation}
with parameters 
\begin{align}
  &q_{1}=26.409 \pm 0.003 \qquad r_{1}=0.087420 \pm 0.000009 \label{eq:fit_x1}\\
  &q_{2}=8.941 \pm 0.002 \qquad \,\,\,r_{2}=0.087116 \pm 0.000007\,, \label{eq:fit_x2}
\end{align}
The good quality of the fit can be assessed from the plot in Fig. \ref{fig:fit_one_domain}. This outcome confirms that the domain motion does not perceptively slow down on this time scale and that the distance between domain walls does not change during the evolution. The space-time dependence of the energy density $\epsilon$ and the transferred momentum density $\langle T^t_{~x} \rangle$ is shown in Fig. \ref{fig:energy_density_and_Txt_one_domain}.
Thus, the friction is too small to be observed at these velocities.
Unfortunately, we encounter severe numerical difficulties when trying to
significantly increase the velocity, so we leave this investigation for the future.


\subsection{Collision of two phase domains}

The solutions of the Einstein-matter equations for a single moving domain at a given time $t^*$ can be used to construct the initial conditions for two domains, each moving
in the opposite direction. In order to represent a system in which a \textit{left} energy domain translates towards positive values of $x$ and a \textit{right} one moves in the opposite direction, the amplitude of the $x$ interval has to be doubled to $24\pi$. Hence, the physics of the right energy domain can be obtained from that of the left one by reflection with respect to the $x=12\pi$ axis: $x\to 24\pi-x$. Upon such coordinate transformation in the line element \eqref{eq:metrictimedep}, only the metric function $B$ (therefore $\tilde{B}$) switches its sign.

The initial conditions for the counterpropagation of two energy domains can thus be obtained by gluing two single-domain solutions as follows:
\begin{align}
\tilde{S}_{2dom}(z,0,x)=& 
\tilde{S}_{1dom}(z,t^*,x) \chi_{[0,12\pi[}(x)+\tilde{S}_{1dom}(z,t^*,24\pi-x)\chi_{[12\pi,24\pi[}(x)\,,\label{eq:S2dom}\\
\tilde{G}_{2dom}(z,0,x)=& 
\tilde{G}_{1dom}(z,t^*,x) \chi_{[0,12\pi[}(x)+\tilde{G}_{1dom}(z,t^*,24\pi-x)\chi_{[12\pi,24\pi[}(x)\,,\label{eq:G2dom}\\
\tilde{B}_{2dom}(z,0,x)=& 
\left(\tilde{B}_{1dom}(z,t^*,x)-\tilde{B}_{1dom}(z,t^*,12\pi)\right) \chi_{[0,12\pi[}(x) \nonumber \\
-& \left(\tilde{B}_{1dom}(z,t^*,24\pi-x)-\tilde{B}_{1dom}(z,t^*,12\pi)\right)\chi_{[12\pi,24\pi[}(x)\,,\label{eq:B2dom}
\end{align}
where $\chi_X(x)$ is the characteristic function of set $X$, and the functions $\left(\tilde{S}_{1dom},\tilde{G}_{1dom},\tilde{B}_{1dom}\right)$ on the right hand sides correspond to the solution of a single moving domain at time $t^*$, with the origin of the $x$ axis shifted so that their derivatives vanish at $x=0,12\pi$. This choice is made to avoid cusps in the junction of the two branches. Moreover, to ensure the continuity of $\tilde{B}_{2dom}$, both branches have been rescaled to obtain the same (vanishing) value at the junction.

\begin{figure}
	\centering
		\includegraphics[width = 0.75\textwidth]{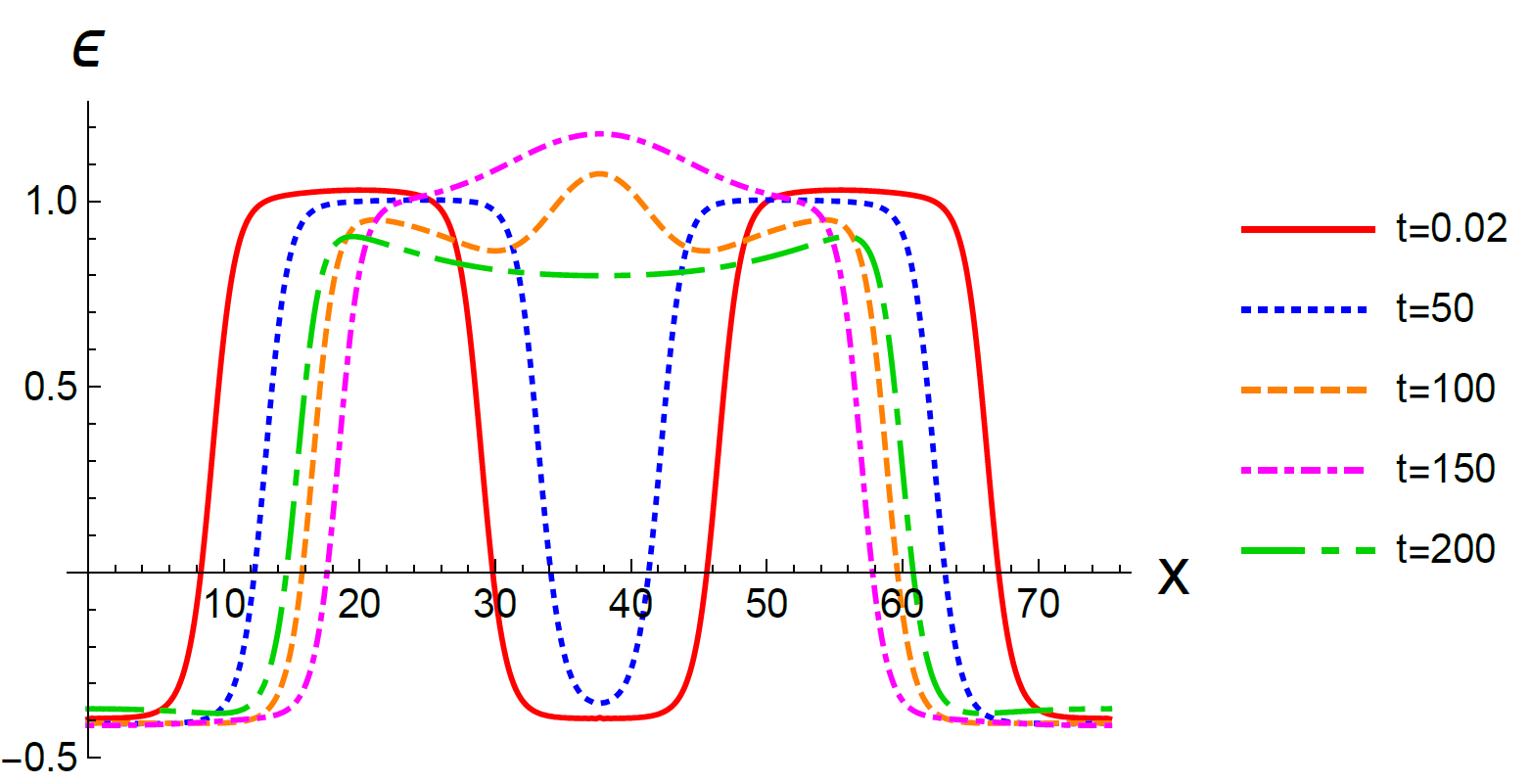}
	\caption{Energy density profile obtained solving the Einstein-matter equations with initial conditions \eqref{eq:S2dom}--\eqref{eq:B2dom}.
	}
	\label{fig:translation_energy_two_domains}
\end{figure}

To determine our initial conditions as in Eqs. \eqref{eq:S2dom}--\eqref{eq:B2dom}, we consider the evolution at time $t^*=100$ of a single moving domain, obtained by shifting the static $\tilde{B}$ function of the final state in Fig. \ref{fig:finalenergy} by $C=0.05$. The results reported in Fig. \ref{fig:translation_energy_two_domains} show that the two high-energy domains, initially counterpropagating along the $x$ axis, remain stuck to each other after the collision. In the overlap region, the spatial profile of the energy density has a nontrivial height, width and shape evolution, until a steady configuration is reached.


Actually, we have noticed that the numerics becomes unstable as the relative velocity of the two domains is increased. However, the preliminary results show that this analysis is worth further investigations, and the method based on initial conditions \eqref{eq:S2dom}--\eqref{eq:B2dom} can represent a promising framework for future research.

\section{Boost invariant dynamics}
\label{sec6}

In the previous sections we considered either evolution from the spinodal branch
or moving and/or colliding domains at $T_c$. A physically very interesting scenario,
motivated by realistic heavy-ion collisions, is boost invariant expansion.
Here the plasma starts off in the high temperature phase, expands and cools
and eventually the temperature falls below the phase transition temperature.
It is thus interesting to study the real time dynamics of such a system\footnote{See also some more involved shock wave collisions which were studied in \cite{Attems:2018gou,Attems:2017zam}}.
Here we will use this setup also to investigate systems in which new effects appear:
i.e. systems of class B, which possess a new dynamical instability
and systems of class C, which exhibit a confinement-deconfinement phase transition
and which do not have a low temperature black hole phase.

Let us adopt a flat Minkowski metric $ds^2=-dt^2+dx^2+d\tilde{y}^2$
and define a coordinate transformation
\begin{equation}
t= \tau \cosh y, \hspace{12pt} \tilde{y} = \tau\sinh y~,
\label{eq:}
\end{equation}
where $\tau$ is the proper boundary time and $y$ is the rapidity.
The inverse transformation has the following form
\bea
y = \frac{1}{2}\log\left(\frac{t+\tilde{y}}{ t-\tilde{y}}\right),\hspace{12pt} \tau =\sqrt{t^2-\tilde{y}^2}~,
\label{eq:transform}
\hspace{12pt}
ds^2 = -d\tau^2 +dx^2+ \tau^2 dy^2 ~.
\label{eq:dsbf}
\eea
In the boundary field theory boost invariance is essentially 
the system's independence on rapidity variable. This reflects
the intuition that at infinite energy nothing depends on finite
boosts. 

The dual geometry admits the following, natural metric ansatz 
\begin{equation}
\label{eq:Metricboost}
ds^2 = -A\,dv^2 - \frac{2\,dv\,dz}{z^2} + S^2\left(G\,dx^2 + v^2\,G^{-1}\,dy^2\right)-2B\,dv\,dx
\end{equation}
where metric functions $A, S, G, B$ and scalar field $\phi$ are functions of radial coordinate $z$, the EF time $v$ (which reduces
to $\tau$ at $z=0$) and  $x$ which is coordinate transverse to the flow. When the system is $x$-independent we have a homogeneous boost invariant flow.
This is particularly important, since the potential class $V_B$ has an instability at $k=0$ which can be seen in a homogeneous time evolution \cite{Gursoy:2016ggq}.


\begin{figure}
	\begin{center}
	\includegraphics[height=.23\textheight]{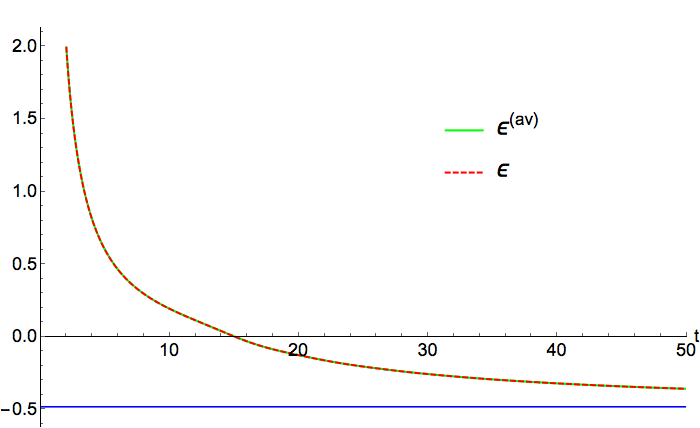}\includegraphics[height=.23\textheight]{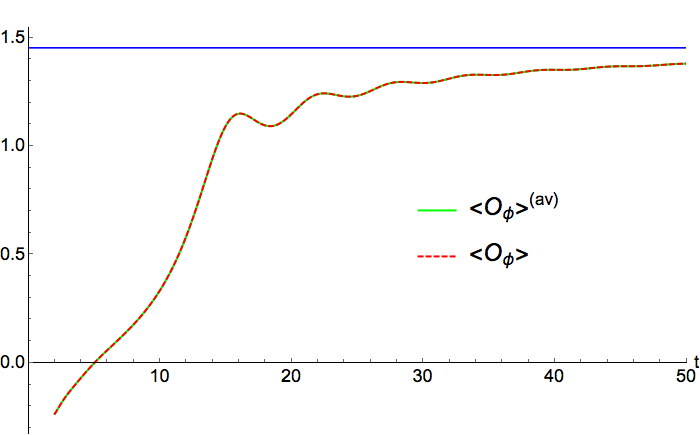}
	\caption{Comparing the mean value of one point functions of inhomogenous evolution with the homogenous one for the class A potential. The blue horizontal lines correspond to the smallest black hole solutions which we expect to be the late time state of the boost invariant expansion.}
	\label{BI-Inhomo-0.2}
	\end{center}
\end{figure}


In order to see the effects of inhomogenity in
a boost invariant evolution we perturb the system
by adding two contributions to the initial geometry
\bea
\delta S(z,x)=S_0\,z^2(1-z)^3\,\cos(x/6) ~, \qquad b_1(x)=B_0\,\sin(x/6) ~,\label{initial-inhomo}
\eea
%
with small amplitudes $S_0$ and $B_0$.
Starting point
for the evolution is $\tau_0=2$. In the inhomogeneous case the
time evolution of one point functions averaged over one spatial
period follows closely the evolution of corresponding homogeneous
one-point functions. Since  the apparent horizon boundary condition is imposed in time evolution routine one may try to understand the behaviour of the evolution in comparison with the black hole equation of state shown in Fig.~\ref{fig:eos3classes}. 
In Fig.~\ref{BI-Inhomo-0.2} we plot the energy density of the boundary theory $\varepsilon$ and the density expectation value of the scalar field for class A potential.
The blue horizontal lines  correspond to the smallest black hole solutions expected to be related to the late time solutions and the time evolution for (in)homogenous expansions confirm this expectation. Note that the average of the energy density  in one period along the x-direction is decreasing in time (as expected) and the 1pt-function of the scalar field is oscillating around a   monotonic function. But surprisingly,   they follow the homogenous time evolution which shows that the inhomogeneous perturbation washes out during the time evolution and hydrodynamic instabilities don't enhance the inhomogenity. In both cases our codes breakdown in quite late time $\tau_*$ when $\frac{\varepsilon(\tau_0)-\varepsilon(\tau_*)}{\varepsilon(\tau_0)-\varepsilon(\infty)}>0.95$. Investigating in our numerical code and comparing with black hole solutions we learned that the breakdowns are due to the low number of grid points along the radial coordinate. If we want to use higher number of grid points we would have to increase the numerical precision of the routine too, which unfortunately is not possible in the Python
framework that we are using while keeping acceptable run time performance.

\begin{figure}
	\begin{center}
	\includegraphics[height=.23\textheight]{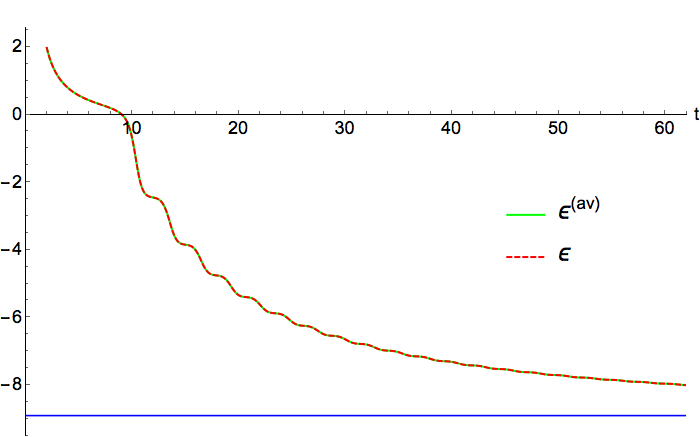}\includegraphics[height=.23\textheight]{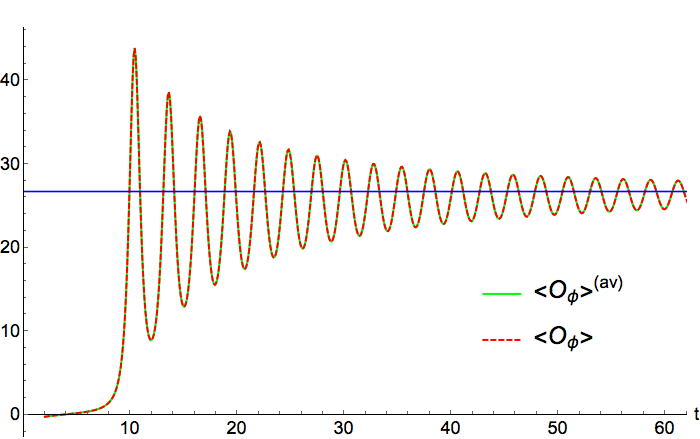}
	\caption{Comparing the mean value of one point functions of inhomogenous evolution with the homogenous one for the class B potential. Again the blue horizontal lines correspond to the smallest black hole solutions which we expect to be the late time state of the boost invariant expansion.}
	\label{BI-Inhomo-0.3}
	\end{center}
\end{figure}

In Fig. \ref{BI-Inhomo-0.3} one can see the results for class B potential. As we have emphasized in section~\ref{sec2}, the main difference between two potentials is the dynamical instability through the first nonhydro QNMs which does exist even at zero momentum (homogeneous evolution). While the average of the  energy density is again monotonically decreasing, as we expect for boost invariant expansion, it is interesting to see the effect of dynamical unstable modes in time evolution of the expectation value of the scalar field in the right panel of Fig. \ref{BI-Inhomo-0.3}. Interestingly, the behaviour shows an exponential growth followed by an oscillation around the expected final value $\langle O_\phi\rangle|_{(v=\infty)}$. For the same reason we explained in the previous paragraph the numerics breakdown in the late time $\tau>\tau_*$ when more than $90\%$ of the relative energy density is reduced $\frac{\varepsilon(\tau_0)-\varepsilon(\tau_*)}{\varepsilon(\tau_0)-\varepsilon(\infty)}>0.9$. Again the average of the 1pt-functions in one period along x-direction follow the homogenous time evolution.


\section{Theories with a confinement-deconfinement phase transition}
\label{sec7}

\begin{figure}
\begin{center}
\includegraphics[height=.22\textheight]{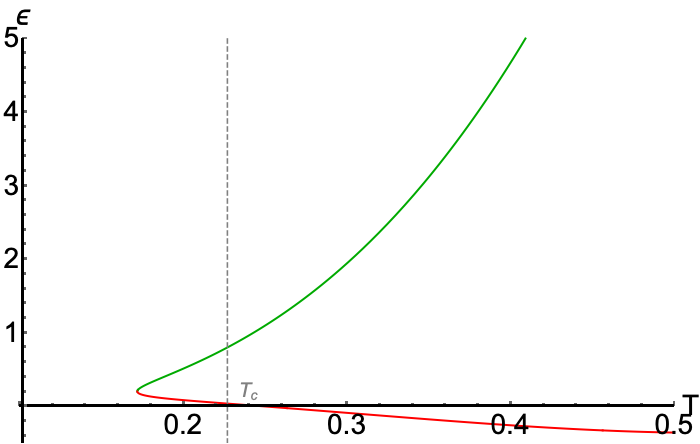}\includegraphics[height=.22\textheight]{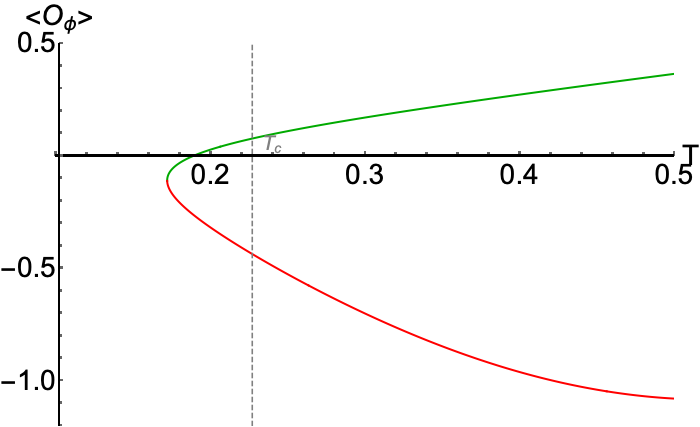}
\caption{The energy (left) and free energy (right) of the dual boundary theory corresponding to the class C potential.}  
\label{1ptfclassC}
\end{center}	
\end{figure}

In this section we focus on  class C potential in table \ref{tabV} which corresponds to the confinement-deconfinement phase transition. 
As  illustrated in lower panels in Fig. \ref{fig:eos3classes} 
and in Fig. \ref{1ptfclassC},
there are two branches of homogenous black hole solutions for temperature higher than $T_{\text min}\sim0.172$. While for lower temperature the only solution to the Einstein equations of motion is a thermal gas at given temperature with zero free energy.  The transition between thermal gas (confinement phase) and homogenous black holes (deconfinement phase) occurs at critical temperature $T_c\sim0.227>T_{\text min}$. 
Therefore, this is a proper setup to study the confinement-deconfinement phase transition which is a transition between two different phases of matter. But from gravity perspective this is rather a difficult task since the topology of black holes (with a horizon) is completely different than a thermal gas (without a horizon).

At this point we would like to bring up another technical problem. Using spectral Chebyshev method has some limits for this potential. In small black hole branch  to find black holes with temperature higher than $T\sim0.35$ (corresponding to $\phi_H\sim5$) large number of grid points (more than 120) and high accuracy are needed. Since there are some limits in the Python packages that we use we restrict our self to 80 grid points in radial coordinate. This leads to numerical inaccuracy and breaking down of the code whenever some local value of the scalar field at the horizon goes larger than critical value $\phi_H\sim5$. On the other hand, the small black with critical temperature $T_c$  corresponds to $\phi_H\simeq3.185$ which guarantees that we are able to investigate the physics near to the critical temperature.

\begin{figure}
\begin{center}
\includegraphics[height=.22\textheight]{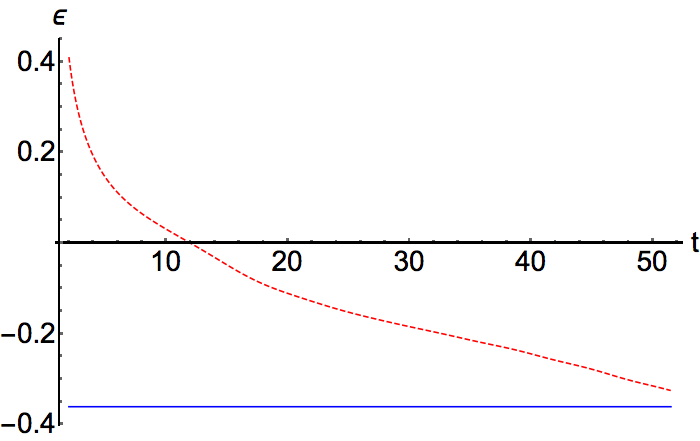}
\includegraphics[height=.22\textheight]{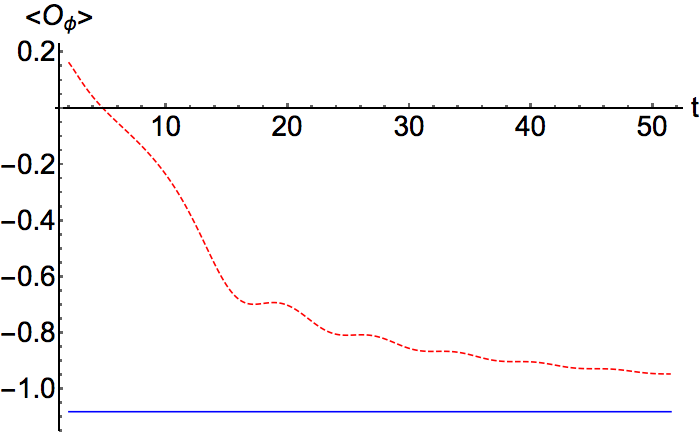}
\caption{The energy (left) and expectation value of the operator dual to the scalar field (right) as a function of time in boost invariant flow for the class C potential. The horizontal blue lines correspond to the black hole with lowest energy density. }  
\label{boosthomoClassC}
\end{center}	
\end{figure}

Although we were not able to study the formation of phase domains in this setup, we investigate the boost invariant expanding plasma under certain circumstances. We impose the apparent horizon boundary condition during the time evolution which forces the plasma to almost follow the equation of state of the static solutions given in the most bottom panel of Fig. \ref{fig:eos3classes} until the plasma  enters the numerically unstable regime of the code. In Fig. \ref{boosthomoClassC} we show the results for homogenous expansion starting from $\phi_H=1$ black hole which lives in the large black branch. By comparing Fig. \ref{1ptfclassC} and \ref{boosthomoClassC}  one can see the critical temperature corresponds to $\varepsilon(T_c)\simeq0.04$ and $\langle O_\phi(T_c) \rangle\simeq-0.43$ in unstable branch. This point is passed around $v\sim 12$ during the expansion and it approaches the black hole with lowest energy density as this is the expected final state of our homogeneous boost invariant expansion. In particular, we do not see 
any consistent breakdown which would indicate a passage in the direction of the stable 
thermal gas background. 

\section{Summary and Outlook}
\label{sec8}

In this paper we carried out a detailed study of the time evolution of a number of holographic  strongly coupled models in 2+1 dimensions undergoing first order phase transition.
This type of models was introduced in \cite{Gubser:2008ny,Gursoy:2007er,Gursoy:2007cb} and includes 3+1 dimensional gravity coupled to a scalar field with a given self-interacting potential which specifies the model.
In classes A and B the phase transition is between two black holes while the third class C, which is more interesting from boundary point of view, exhibits a transition between a black hole and thermal gas.

The real time dynamics of the boundary theory in strongly coupled regime can be investigated by solving the classical equations of motion in the dual gravity theory for an out of equilibrium initial configuration.
We have listed several open questions related to this setup and, in our specific models, we have performed an extensive study of the time evolution of their spinodal instabilities. 
Our main observations are following.

Firstly, we identified a couple of common features in the final state starting from the spinodal branch in class A potential for various perturbations. We observed a pattern of merging of phase domains. These occurred when the domain in between was either very narrow or the merging happened through collisions. Wider, static phase domains were extremely long lived. Furthermore, we verified that static phase domains have an exponentially long lifetime. Thus the landscape of final states 
contains a variety of inhomogeneous black holes which, for all practical purposes,
have (at least) an exponentially long lifetime.

Unfortunately we could not repeat the same investigation for other classes of theories because of some numerical instabilities appearing in our approach. While the study for class B potential may need more accurate calculation (higher number of grid points which needs stronger computers than normal desktop/laptop that we have used), class C seems to be more challenging due to different topology of the black hole and thermal gas in two phases\footnote{For an interesting investigation of the confinement-deconfinement transition and relevant technical difficulties see \cite{Hanada:2018zxn}.}.
Further study of these two models are still open tasks to be done in future.

Secondly, apart from observing the merging of domains within the extended simulation from the small perturbation until the final state, we investigated the possibility of
studying directly collisions of fully formed moving phase domains by constructing
appropriate initial conditions. This was done by first making a moving domain in one period and then gluing with its mirror image moving in the opposite direction.

Thirdly, we have used the boost invariant setup to learn what would be the effect of spinodal instability on an expanding plasma.
To this end we compared the results for homogeneous and inhomogeneous expansions for class A and class B potentials.
Our results show that while the non-hydro instability clearly manifests itself in the comparison between two models, the inhomogeneity washes out when the energy density passes the spinodal instability.
We also set the same calculation for class C potential with homogeneous expansion.
Since in this setup we impose the apparent horizon boundary condition the evolution is effectively following the equation of state of homogeneous black holes, showing no sign
of transition to the thermal gas phase.

We close this section by listing some open questions.
While the phase transition between a black hole and a thermal gas is physically most interesting, it is the most challenging one as well, and as such needs further studies.
One may expect, that purely classical gravity may not suffice in this case.
The theories with dynamical (nonhydrodynamic) instability are easier, but still
require significantly larger numerical resources. An interesting further avenue of research would be to pursue the study of collisions of moving phase domains in more detail and understanding the difficulties in constructing moving phase domains with higher velocity.

While this paper was in the final stages of completion, an interesting work \cite{Attems:2019yqn} appeared, which shares some similar results with the present paper,
but works in the context of a 4+1 dimensional gravity coupled to a self interacting scalar field.


\ackno
LB was supported by the Angelo Della Riccia Foundation and the Jagiellonian University during her stay at the Marian Smoluchowski Institute of Physics in Krakow, where part of this research activity was carried out.
 JJ and HS would like to thank Jagiellonian University, for its hospitality during various visits.
 JJ  was  supported by the by the Polish National Science Centre (NCN)  grant 2016/23/D/ST2/03125. RJ was supported by NCN grant 2012/06/A/ST2/00396.

\appendix

\section{Details on the numerical procedure}
\label{appA}

In this appendix we provide some technical details on the
procedure of the numerical evolution for the geometries given by \eqref{eq:metrictimedep} and \eqref{eq:Metricboost}.  The advantage of using the following ansatz in Eddington-Finkelstein coordinates is twofold. Firstly, it encompasses both the standard and the boost-invariant time evolution.
Secondly, the resulting numerical calculations are rather stable.
The ansatz for the metric is given by
\be
ds^2=-A\,dv^2-\frac{2\,dv\,dz}{z^2}-2\,B\, dv\,dx + S^2\,\left( G\,dx^2+f(v)\,G^{-1}\, dy^2 \right),
\label{eq:Metric}
\ee
where $A, B, S, G$ are functions of $z, v, x$ and auxiliary function $f(v):=c_1+c_2 v^2$ is defined such that for the fixed energy studies $(c_1, c_2)=(1, 0)$ and for the boost invariant expansion $(c_1, c_2)=(0, 1)$. Note that the coordinates $v, y$ in this ansatz are different in each case.

In general, dealing with the time evolution, we will follow the strategy reviewed in \cite{Chesler:2013lia} and define
\be
d_+:=\partial_v-\frac{z^2 A}{2}\partial_z~,\label{dplus}
\ee
The coupled set of Einstein-matter equations read
\bea
&&\phi _{,z}-\left(-\frac{(G_{,z})^2}{G^2}-\frac{8S_{,z}}{z \,S}-\frac{4S_{,z^2}}{S}\right)^{\frac{1}{2}}=0\label{inhomo-eq1}\\
&&B_{,z^2}+B_{,z} \left(\frac{2G-z G_{,z}}{z G}\right)+\frac{B}{2} \left(\frac{3 (G_{,z})^2}{G^2}-\frac{4 (S_{,z})^2}{S^2}+\phi _{,z^2}-\frac{4G_{,z} (\frac{1}{z}+\frac{S_{,z}}{S})+2G_{,z^2}}{G}\right)\qquad\nn\\
&&\qquad=R_{B}(G, S, \phi, f)\label{inhomo-eq2}\\ 
&&(d_+ S)_{,z}+\frac{d_+ S S_{,z}}{S}=R_{d_+S}(G, S, \phi, B, f)\label{inhomo-eq3}\\ 
&&(d_+ G)_{,z}+d_+ G \, \left(\frac{S_{,z}}{S}-\frac{G_{,z}}{G}\right)=R_{d_+G}(G, S, \phi, B, d_+S, f)\label{inhomo-eq4}\\
&&d_+ \phi _{,z}+\frac{(d_+\phi) S_{,z}}{S}=R_{d_+\phi}(G, S, \phi, B, d_+S, f)\label{inhomo-eq5}\\ 
&&A_{,z^2}+\frac{2 A_{,z}}{z}=R_{A}(G, S, \phi, B, d_+S, d_+G, d_+\phi, f)\label{inhomo-eq6}\\ 
 &&(d_+B)_{,z}-d_+B \left(\frac{G_{,z}}{G}+\frac{2 S_{,z}}{S}\right)=R_{d_+B}(G, S, \phi, B, d_+S, d_+G, d_+\phi, A, f)\label{inhomo-eq7}\\ 
&&d_+^2 S=R_{d_+^2S}(G, S, \phi, B, d_+S, d_+G, d_+\phi, A, d_+B, f)\label{inhomo-eq8}
\eea
with source terms given in the right hand side of the equations.
In the above formulas 
$\mathcal{A}_{,z}:=\partial_z\mathcal{A},~\mathcal{A}_{,z^2}:=\partial_z^2\mathcal{A}$. Motivated by the near boundary solutions and numerical convenience we introduce following re-definitions 
\bea
&&\tilde{\phi}:=\frac{\phi}{z}=\varphi_1+\varphi_2\, z+\mathcal{O}(z^2)~,\label{asym-inhomo-1}\\
&&\tilde{S}:=z S=1+s_1\,z+\mathcal{O}(z^2),\label{asym-inhomo-2}\\ 
&&\tilde{A}:=\frac{z^2A-1}{z}=2 s_1-\frac{\dot{f}}{2f}-\left(\frac{\varphi_1^2}{4}-s_1^2+2\dot{s_1}+\frac{s_1 {\dot{f}}}{2\,f}+\frac{3{\dot{f}}^2}{16f^2} \right)z+a_3z^2+\mathcal{O}(z^3)~,\label{asym-inhomo-3}\\ 
&&\tilde{G}:=\frac{G-1}{z}=-\frac{\dot{f}}{2f}+\frac{\dot{f}(4s_1f+\dot{f})}{8f^2}\,z+g_3\, z^2+\mathcal{O}(z^3)~,\label{asym-inhomo-4}\\ 
&&\tilde{B}:=\frac{B+s_1'}{z}=b_1+\mathcal{O}(z)~\label{asym-inhomo-5},\\
&&\wt{d_+S}:=z^2\, d_+S=\frac{1}{2}+\mathcal{O}(z)~\label{asym-inhomo-6},\\
&&\wt{d_+G}:=\frac{d_+G-\dot{f}/(4f)}{z}=\frac{\dot{f}^2-2\,f\,\ddot{f}}{4\,f^2}+\mathcal{O}(z)~\label{asym-inhomo-7},\\
&&\wt{d_+\phi}:=\frac{d_+\phi+{\varphi_1}/2}{z}=-\varphi_2-{\varphi_1}s_1{+\dot{\varphi_1}}+\frac{\dot{f}}{4\,f}+\mathcal{O}(z)~\label{asym-inhomo-8},
\eea
where $\dot{\mathcal{A}}:=\partial_v \mathcal{A}, {\mathcal{A}}':=\partial_x \mathcal{A}$. Since we are interested in fixed source value for the scalar field we choose $\varphi_1=1$.
The unknown functions $s_1, a_3, g_3, b_1$ and $\varphi_2$ are related to the 1-point functions of the dual theory and they can not be fixed by near boundary analysis. Nevertheless, by imposing the asymptotic boundary conditions and the ones at the apparent horizon, and solving the Einstein-matter equations of motion one can find them.
The near boundary analysis also shows two extra equations related to the boundary Ward identities,
\bea
\dot{a_3}&=&-\frac{\dot{s_1} + \dot{\varphi_2} -3 {b_1}'}{2}-\frac{\dot{f}\,\left(3 a_3 + 3 g_3 + f_2 \right)}{4\,f}-\frac{s_1}{8\,f^3}\left( 3 s_1 f \, \dot{f}^2+{4 f^2 \dot{f}+3 \dot{f}^3} \right)\nn\\
&&+\frac{13 {\dot{f}}^4}{128 f^4}-\frac{3 {\dot{f}}^2}{32 f^2}+\left(\frac{1}{8 f}-\frac{{\dot{f}}^2}{4 f^3}\right) \ddot{f} ~,\label{ward1}\\
\dot{b_1}&=&\frac{{a_3}'-3 {g_3}'}{3}+\frac{\dot{f}}{f} \left(\frac{ b_1}{2}+{s_1 {s_1}'}+\frac{\dot{f} {s_1}'}{4\,f}\right)~.\label{ward2}
\eea
which reflect the Ward identity  and energy-momentum conservation which are explicitly driven in next appendix \ref{appB}.

\subsection{Numerical routine}

In all  simulations performed in this paper we follow a generic strategy to solve equations of motion originating
in the characteristic formulation reviewed in \cite{Chesler:2013lia}.
We use Chebyshev pseudo spectral discretization  along the radial coordinate with the number of grid points between 50 and 100. 
The plots presented in this paper are with 80 number of grid points. 
For integration along the spatial field theory direction $x$ we use Fourier transformation with large enough number of grid points adjusted for each period. 
Also to find stable results, we implemented so called sharp low-pass filtering method with keeping  the first sixty percent of the Fourier modes at each time step and setting to zero all higher modes\footnote{For the simulations for the merging domains
we employed also Chebyshev filtering as well as filtering the resulting functions
every $\Delta t=0.01$.}.
To be specific, the procedure consists of the following steps
\begin{enumerate}
\item At $v=v_0$, knowing functions $S, G$ and one boundary condition for the scalar field which is $\wt{\phi}(0, v, x)=1$, one can integrate Eq. \eqref{inhomo-eq1} to find $\phi(z,v_0,x)$. We are interested in a setup in which  the $S$ function of a static solution has been perturbed by
\bea
\delta\tilde{S}_{\rm pert}=S_0\,e^{\sigma (z-1/2)^2}\,z^3\,(1-z)^3\,\sigma_1(x)~.
\eea
Note that this is a perturbation for redefined function $\tilde{S}$ which differs by a factor of $z$ with function $S$ in the original metric ansatz.
The initial configuration is given by choosing the values for the parameters $S_0$ and the appropriate function $\sigma_1(x)$ depends on specific questions. Various examples are considered in the remainder of the paper. For boost invariant setup we also modify the near boundary of the $S$ and $G$ functions according to their time dependency.
\item Linearly independent asymptotic solutions of $B$ function in \eqref{inhomo-eq2} behave as $z^{-2}$ and $z^1$. The coefficient of the $z^{-2}$ homogeneous solution has to be zero and we just need to fix $b_1(v_0,x)$. We start with
\be
b_1(v_0,x)=\sigma_2(x)~,
\ee 
Again $\sigma_2(x)$ depends on specific questions we are interested in. 
\item We can find $\wt{d_+S}$ by solving \eqref{inhomo-eq3} with one boundary condition,
\be
\wt{d_+S}(z_H,v_0,x)=\bigg[\frac{1}{2\,G\,S}\left(\frac{z^2\, B^2\, \rd_z S}{S}+\frac{B\,\rd_x G}{G}-{\rd_x B}\right)-\frac{\dot{f}\,S}{4\,f}\bigg]_{z=z_H}~,\label{dpS-Horizon-bc}
\ee
which comes from the definition for apparent horizon \cite{Chesler:2013lia}.
\item Using  \eqref{inhomo-eq4} and one boundary condition, $\wt{d_+G}(0,v_0,x)={(\dot{f}^2-2\,f\,\ddot{f})}/{(4\,f^2)}$, one can find $\wt{d_+G}$. 
\item Using  \eqref{inhomo-eq5} and one boundary condition, $\wt{d_+\phi}(0,v_0,x)=-\varphi_2-s_1+\frac{\dot{f}}{4\,f}$, one can find $\wt{d_+\phi}$.
\item To find $\wt{A}$ we  should solve   \eqref{inhomo-eq6} with two boundary conditions. The first one is $\wt{A}(0,v_0,x)=2\,s_1-\frac{\dot{f}}{2f}$ but the second one is related to the stationary horizon condition. 
Using equations \eqref{inhomo-eq8} and \eqref{dpS-Horizon-bc} one can find the corresponding second order elliptic equation for $\wt{A}$ at the horizon which can be solved in terms of Fourier modes.
\item By having functions $S, G, \phi, A, d_+S, d_+G, d_+\phi$ and using the definition \eqref{dplus} one can go one step forward in time $v=v_0+\delta v$ and repeat the routine from the first step. For integration in time we use fourth-order Runge-Kutta  method for the first three time steps and then we use the fourth order Adams-Bashforth method. 
Note that to find $b_1(v_0+\delta v,x)$ in next time step we use the Ward identity \eqref{ward2} which is founded from the near boundary expansion of \eqref{inhomo-eq7}. 
So, in this routine we are using all the equations of motion.
\end{enumerate}
\section{Holographic renormalization}\label{appB}

In this section by applying the Hamilton-Jacobi formalism \cite{Elvang:2016tzz} we will find the counter-terms and will calculate the boundary energy-momentum tensor. It is more convenient for our propose to use Fefferman-Graham (FG) coordinates,
\bea
ds^2=dr^2+\gamma_{ij}\, dx^i \, dx^j~.
\eea
The divergent part of the counterterm action is
\be
S_{\rm ct}=-\frac{1}{\kappa^2}\int_{\partial M}d^3x \, \sqrt{\gamma}\, \left( \mathcal{U}_0+\mathcal{U}_2 \right)~,
\ee
where $\mathcal{U}_i$ has $i$ derivatives. The gravity part of these terms is known  \cite{Elvang:2016tzz}.
To find the scalar field contribution we need to include terms which are potentially divergent at $\mathcal{U}_i$ with proper derivatives and solve the Hamiltion-Jacobi equation,
\be
R[\gamma]+\mathcal{K}+2\,p^2-\frac{1}{2}\gamma^{ij} \, \partial_i \phi \, \partial_j \phi-V(\phi)+2 \, \partial_r \left(  \mathcal{U}_0+\mathcal{U}_2 \right)=0~,
\ee 
order by order in derivatives where $\mathcal{K}=K_{ij}K^{ij}-K^2$ is the extrinsic curvature contribution. Since the scalar field has asymptotically $e^{-r/L}$ falloff and the action is invariant under $\phi \rightarrow -\phi$, the most general Ansatz for zero-derivative order is
\bea
\mathcal{U}_0=-\frac{2}{L}+\alpha(r) \, \phi^2~.
\eea
Keeping only 0-derivative terms and using $\mathcal{K}_0=-\frac{3}{2}\mathcal{U}_0^2$ one can find
\bea
-\frac{3}{2}\mathcal{U}_0^2+2\,p_0^2-V(\phi)+2\, \partial_r\, \mathcal{U}_0=0~,
\eea
where
\be
p_0=\frac{\partial \mathcal{U}_0}{\partial \phi}=2 \, \alpha\, \phi~.
\ee
By comparing this expansion with the definition 
\be
p=\frac{\kappa^2}{\sqrt{\gamma}}\frac{\partial\, S}{\partial \phi'}=-\frac{1}{2L}\phi+\mathcal{O}(e^{-2r/L})
\ee
one can see that $\alpha=-1/(4L)$ and the 0-derivative term is
\be
\mathcal{U}_0=-\frac{2}{L}-\frac{1}{4L}\, \phi^2~.
\ee

At two-derivative order there is no contribution from the scalar field because of the symmetries,
\bea
\mathcal{U}_2=-\frac{L}{2}\,R[\gamma]~.
\eea

Using the counterterms and the regularized on-shell action we can 
obtain the renormalized generating functional
\begin{equation}
S_{\rm ren} = \lim_{r\rightarrow\infty}S_{\rm reg}=\lim_{r\rightarrow\infty}(S_{\rm on-shell} + S_{\rm ct})~.
\label{eq:}
\end{equation}
Our favourite one point functions are given as functional derivatives
of the  generating functional i.e.
\bea
\langle {O_{\phi}} \rangle &=& \lim_{r\rightarrow\infty} \frac{e^{\Delta r}}{\sqrt{-\gamma}}\frac{\delta S_{\rm reg}}{\delta\phi}=
\frac{1}{2}\lim_{r\rightarrow\infty}e^{\Delta r}(\partial_r\phi+\phi)~,\\
\langle T_{ij} \rangle &=&2 \lim_{r\rightarrow\infty} \frac{e^{r}}{\sqrt{-\gamma}}\frac{\delta S_{\rm reg}}{\delta\gamma^{ij}}\\
&=&2\lim_{r\rightarrow\infty}e^{r}\Bigg[-\frac{1}{4}(\partial_r \gamma_{ij}-\gamma_{ij} \gamma^{kl}\partial_r\gamma_{kl})+\frac{1}{2}R_{ij}[\gamma]-\frac{1}{4}\gamma_{ij}(R[\gamma]+1)-\frac{1}{8}\gamma_{ij}\phi^2\Bigg]~,\nn
\eea
from which, after direct evaluation and hiring the coordinate transformation from FG to EF ansatz, one gets
\begin{equation}
\langle O_\phi\rangle =- \frac{1}{2}(s_1+\varphi_2)+\frac{\dot{f}}{8\,f}  ~.
\label{eq:}
\end{equation}
and the energy-momentum tensor yields the following shape,
\bea
&&\varepsilon:=T_{t}^{t} = -a_3 -\frac{1}{2}(s_1+\varphi_2)+\frac{\dot{f}}{8\,f} ~,\\
&&P_x:=T_{~x}^x = \frac{a_3}{2}-\frac{3}{2} g_3-\frac{\dot{f}}{4\,f}\,\left( 3\,{s_1}^2+\frac{3\,s_1 \dot{f}}{2\,f} +\frac{4\,f^2-13\,\dot{f}^2+32\,f\,\ddot{f}}{16\,f^2}  \right), \\
&&P_y:=T_{~y}^y = \frac{a_3}{2}+\frac{3}{2} g_3+\frac{\dot{f}}{4\,f}\,\left( 3\,{s_1}^2+\frac{3\,s_1 \dot{f}}{2\,f} +\frac{4\,f^2-13\,\dot{f}^2+32\,f\,\ddot{f}}{16\,f^2}  \right), \\
&&T_{~x}^{t} =-\frac{3\,b_1}{2}.
\eea
in terms of the near boundary data in EF coordinate.

It is straightforward to see the above stress tensor satisfies the known Ward identities,
\be
\nabla_i \langle T^{i j} \rangle=(\partial^j\varphi_1)\,\langle {O_\phi} \rangle=0,\qquad\qquad\langle {T_i}^i \rangle=\varphi_1 \langle {O_\phi}\rangle=\langle {O_\phi}\rangle~.
\ee
The second equality in both equations reduce the results in our setup since we are interested in the cases with unit value for the source of the scalar field.

\bibliography{bibdynam}
\bibliographystyle{bibstyl}

\end{document}